\documentclass[12pt]{article}
\usepackage[dvips]{epsfig}
\usepackage{amsmath,amssymb,amsfonts,enumerate,bbm}
\usepackage{caption}
\usepackage{url}
\usepackage{ifthen}
\usepackage{graphicx}

\usepackage{subfigure}
\usepackage[pdftex]{hyperref}
\usepackage{color,soul}

\newcommand{\Ot}{\tilde{O}}

\long\def\commabs #1\commabsend{}
\long\def\commful #1\commfulend{}
\long\def\comment #1\commentend{}

\newtheorem{theorem}{Theorem}[section]
\newtheorem{lemma}[theorem]{Lemma}
\newtheorem{observation}[theorem]{Observation}
\newtheorem{corollary}[theorem]{Corollary}

\newtheorem{definition}[theorem]{Definition}

\newcommand{\REAL}{\mathbb R}
\def\deg{\mbox{\tt deg}}
\def\depth{\mbox{\tt depth}}

\def\Set{\mathfrak{F}}

\def\Cost{\mbox{\tt Cost}}

\newcommand{\dist}{\mbox{\rm dist}}

\def\inline#1:{\par\vskip 7pt\noindent{\bf #1:}\hskip 10pt}
\def\Proof{\par\noindent{\bf Proof:~}}
\def\blackslug{\hbox{\hskip 1pt \vrule width 4pt height 8pt
    depth 1.5pt \hskip 1pt}}
\def\QED{\quad\blackslug\lower 8.5pt\null\par}

\newcommand{\cS}[0]{{\cal S}}

\newcommand{\New}[0]{\mbox{\tt New}}

\def\LastE{\mbox{\tt LastE}}

\def\NSource{\sigma}
\def\Root{\mbox{\tt r}}
\def\Leaf{\mbox{\tt Leaf}}
\def\NLeaf{\mbox{\tt nLeaf}}

\def\Depth{\mbox{\tt Depth}}

\def\FTMBFS{\mbox{\tt FT-MBFS}}

\def\FTBFS{\mbox{\tt FT-BFS}}
\def\ApproxSetCover{\mbox{\tt ApproxSetCover}}

\begin{document}


\title{Sparse Fault-Tolerant BFS Trees}
\author{
Merav Parter
\thanks{The Weizmann Institute of Science, Rehovot, Israel.
Email: {\tt \{merav.parter,david.peleg\}@ weizmann.ac.il}.
Supported in part by the Israel Science Foundation
(grant 894/09), the United States-Israel Binational Science Foundation
(grant 2008348), the Israel Ministry of Science and Technology
(infrastructures grant), and the Citi Foundation.}
\thanks{Recipient of the Google Europe Fellowship in distributed computing;
 research supported in part by this Google Fellowship.}
\and
David Peleg $^*$
}

\maketitle

\begin{abstract}
A {\em fault-tolerant} structure for a network is required to continue
functioning following the failure of some of the network's edges or vertices.
This paper considers {\em breadth-first search (BFS)} spanning trees,
and addresses the problem of designing a sparse {\em fault-tolerant} BFS tree,
or {\em FT-BFS tree} for short, namely, a sparse subgraph $T$ of the given network $G$
such that subsequent to the failure of a single edge or vertex,
the surviving part $T'$ of $T$ still contains a BFS spanning tree for
(the surviving part of) $G$. For a source node $s$, a target node $t$ and an edge $e\in G$, the shortest $s-t$ path $P_{s,t,e}$ that does not go through $e$  is known as a \emph{replacement path}. Thus, our \FTBFS\ tree contains the collection of all replacement paths $P_{s,t,e}$ for every $t \in V(G)$ and every failed edge $e \in E(G)$.
\par
Our main results are as follows. We present an algorithm that
for every $n$-vertex graph $G$ and source node $s$ constructs
a (single edge failure) \FTBFS\ tree rooted at $s$ with
$O(n \cdot \min\{\Depth(s), \sqrt{n}\})$ edges,
where $\Depth(s)$ is the depth of the BFS tree rooted at $s$.
This result is complemented by a matching lower bound, showing that
there exist $n$-vertex graphs with a source node $s$ for which any edge
(or vertex) \FTBFS\ tree rooted at $s$ has $\Omega(n^{3/2})$ edges.
\par
We then consider {\em fault-tolerant multi-source BFS trees},
or {\em \FTMBFS\ trees} for short,
aiming to provide (following a failure) a BFS tree rooted at each source
$s\in S$ for some subset of sources $S\subseteq V$.
Again, tight bounds are provided,
showing that there exists a poly-time algorithm that
for every $n$-vertex graph and source set $S \subseteq V$ of size $\NSource$
constructs  a (single failure) \FTMBFS\ tree
$T^*(S)$ from each source $s_i \in S$, with $O(\sqrt{\NSource} \cdot n^{3/2})$
edges, and on the other hand there exist $n$-vertex graphs with source sets
$S \subseteq V$ of cardinality $\NSource$, on which any \FTMBFS\
tree from $S$ has $\Omega(\sqrt{\NSource}\cdot n^{3/2})$ edges.
\par
Finally, we propose an $O(\log n)$ approximation algorithm for constructing
\FTBFS\ and \FTMBFS\ structures. The latter is complemented by a hardness
result stating that there exists no $\Omega(\log n)$ approximation algorithm
for these problems under standard complexity assumptions. In comparison with the randomized $\FTBFS$ construction implicit in \cite{GW12},
our algorithm is deterministic  and may improve the number of edges
by a factor of up to $\sqrt{n}$ for some instances.
All our algorithms can be extended to deal with one \emph{vertex} failure as well, with the same performance.
\end{abstract}

\section{Introduction}
\paragraph{Background and motivation}
Modern day communication networks support a variety of logical structures
and services, and depend on their undisrupted operation.
As the vertices and edges of the network may occasionally fail or malfunction,
it is desirable to make those structures robust against failures.
Indeed, the problem of designing fault-tolerant constructions for various
network structures and services has received considerable attention
over the years.

Fault-resilience can be introduced into the network in several different ways.
This paper focuses on a notion of fault-tolerance whereby the structure at hand
is augmented or ``reinforced'' (by adding to it various components)
so that subsequent to the failure of some of the network's vertices or edges,
the surviving part of the structure is still operational.
As this reinforcement carries certain costs, it is desirable to minimize the number of added components.

To illustrate this type of fault tolerance,
let us consider the structure of graph $k$-spanners
(cf. \cite{Peleg00:book,PelegS-89,PelegU-89}).
A graph spanner $H$ can be thought of as a skeleton structure
that generalizes the concept of spanning trees and allows us
to faithfully represent the underlying network using few edges,
in the sense that for any two vertices of the network, the distance
in the spanner is stretched by only a small factor.
More formally, consider a weighted graph $G$ and let $k \geq 1$ be an integer.
Let $\dist(u,v,G)$ denote the (weighted) distance between $u$ and $v$ in $G$.
Then a $k$-spanner $H$ satisfies that
$\dist(u,v,H) \leq  k\cdot \dist(u,v,G)$ for every $u,v\in V$.
\par Towards introducing fault tolerance, we say that a subgraph $H$ is
an {\em $f$-edge fault-tolerant $k$-spanner} of $G$ if
$\dist(u,v,H\setminus F) \leq  k\cdot \dist(u,v,G\setminus F)$
for any set $F\subseteq E$ of size at most $f$,
and any pair of vertices $u,v \in V$. A similar definition applies to
$f$-vertex fault-tolerant $k$-spanners. Sparse fault-tolerant spanner
constructions were presented in \cite{CLPR09-span,DK11}.

This paper considers {\em breadth-first search (BFS)} spanning trees,
and addresses the problem of designing {\em fault-tolerant} BFS trees,
or {\em \FTBFS}\ trees for short.
By this we mean a subgraph $T$ of the given network $G$,
such that subsequent to the failure of some
of the vertices or edges,
the surviving part $T'$ of $T$ still contains a BFS spanning tree for
the surviving part of $G$.
We also consider a generalized structure referred to as a
{\em fault-tolerant multi-source BFS tree}, or {\em \FTMBFS\ tree} for short,
aiming to provide a BFS tree rooted at each source $s\in S$ for some subset
of sources $S\subseteq V$.

The notion of {\em \FTBFS}\ trees is closely related to the problem of
constructing \emph{replacement paths} and in particular to its
{\em single source} variant,
the \emph{single-source replacement paths} problem,
studied in \cite{GW12}. That problem requires to compute the collection
$\mathcal{P}_{s}$ of all $s-t$ replacement paths $P_{s,t,e}$ for every $t \in V$
and every failed edge $e$ that appears on the $s-t$ shortest-path in $G$.
The vast literature on \emph{replacement paths} (cf. \cite{BK09,GW12,RTREP05,TZ05,WY10}) focuses on \emph{time-efficient} computation of the these paths as well as their efficient maintenance in data structures (a.k.a {\em distance oracles}).
In contrast, the main concern in the current paper is with optimizing
the \emph{size} of the resulting fault tolerant structure that contains the
collection $\mathcal{P}_{s}$ of all replacement paths given a source node $s$.
A typical motivation for such a setting is where the graph edges represent
the channels of a communication network, and the system designer would like
to purchase or lease a minimal collection of channels
(i.e., a subgraph $G' \subseteq G$) that maintains its functionality
as a ``BFS tree" with respect to the source $s$ upon any single
edge or vertex failure in $G$.
In such a context, the cost of computation at the preprocessing stage may
often be negligible compared to the purchasing/leasing cost of the resulting
structure. Hence, our key cost measure in this paper is the \emph{size} of
the fault tolerant structure, and our main goal is to achieve {\em sparse}
(or {\em compact}) structures.

Most previous work on sparse / compact fault-tolerant structures and services
concerned structures that are {\em distance-preserving}
(i.e., dealing with distances, shortest paths or shortest routes),
{\em global} (i.e., centered on ``all-pairs'' variants),
and {\em approximate} (i.e., settling for near optimal distances),
such as {\em spanners}, {\em distance oracles} and
{\em compact routing schemes}.
The problem considered here, namely, the construction of \FTBFS\ trees,
still concerns a distance preserving structure.
However, it deviates from tradition with respect to the two other features,
namely, it concerns a ``single source'' variant,
and it insists on exact shortest paths.
Hence our problem is on the one hand easier, yet on the other hand harder,
than previously studied ones.
Noting that in previous studies, the ``cost'' of adding fault-tolerance
(in the relevant complexity measure) was often low (e.g., merely polylogarithmic
in the graph size $n$),
one might be tempted to conjecture that a similar phenomenon may reveal itself
in our problem as well.
Perhaps surprisingly, it turns out that our insistence on exact distances
plays a dominant role and makes the problem significantly harder,
outweighing our willingness to settle for a ``single source'' solution.

\paragraph{Contributions}
We obtain the following results.
In Sec. \ref{sec:prelim}, we define the \emph{Minimum \FTBFS} and
\emph{Minimum \FTMBFS} problems, aiming at finding the minimum such
structures tolerant against a single edge or vertex fault.
Section \ref{sc:lb} presents lower bound constructions for these problems.
For the single source case, in Subsec. \ref{sc:lb_single}, we present a lower bound stating that for every $n$ there exists an $n$-vertex
graph and a source node $s \subseteq V$ for which any
\FTMBFS\ tree from $s$ requires $\Omega(n^{3/2})$ edges.
In Subsec. \ref{sec:multiedge_lb},
we then show that there exist
$n$-vertex graphs with source sets $S \subseteq V$ of size $\NSource$,
on which any \FTMBFS\ tree from the source set $S$ has
$\Omega(\sqrt{\NSource}\cdot n^{3/2})$ edges.

These results are complemented by matching upper bounds.
In Subsec. \ref{subsec:1edge_alg}, we present a simple algorithm that for every
$n$-vertex graph $G$ and source node $s$, constructs a (single edge failure)
\FTBFS\ tree rooted at $s$ with $O(n \cdot \min\{\Depth(s), \sqrt{n}\})$ edges.
A similar algorithm yields an \FTBFS\ tree tolerant to one vertex failure,
with the same size bound.
In addition, for the multi source case, in Subsec. \ref{subsec:multisource}, we show that there exists a polynomial
time algorithm that for every $n$-vertex graph and source set $S \subseteq V$
of size $|S|=\NSource$ constructs  a (single failure) \FTMBFS\ tree $T^*(S)$
from each source $s_i \in S$, with $O(\sqrt{\NSource} \cdot n^{3/2})$ edges.

In Sec. \ref{sec:hardness}, we show that the minimum \FTBFS\ problem is NP-hard and moreover, cannot be approximated
(under standard complexity assumptions) to within a factor of $\Omega(\log n)$,
where $n$ is the number of vertices of the input graph $G$.
Note that while the algorithms of Sec. \ref{sec:upper} match the worst-case lower bounds, they might still be far from optimal for certain instances, as illustrated
in Sec. \ref{sec:approx}.  Consequently, in Sec. \ref{sec:approx}, we complete the upper bound analysis by presenting an $O(\log n)$ approximation algorithm for the Minimum \FTMBFS\ problem.
This approximation algorithm is superior in instances where the graph enjoys
a sparse \FTMBFS\ tree, hence paying $O(n^{3/2})$ edges
(as does the algorithm of Sec. \ref{sec:upper}) is wasteful.
In light of the hardness result for these problems (of Sec. \ref{sec:hardness}),
the approximability result is tight (up to constants).

\paragraph{Related work}
To the best of our knowledge, this paper is the first to study
the sparsity of fault-tolerant BFS structures for graphs.
The question of whether it is possible to construct a sparse fault tolerant
\emph{spanner} for an arbitrary undirected weighted graph,
raised in \cite{CZ03},
was answered in the affirmative in \cite{CLPR09-span}, presenting algorithms
for constructing an $f$-vertex fault tolerant $(2k-1)$-spanner
of size $O(f^2 k^{f+1} \cdot n^{1+1/k}\log^{1-1/k}n)$ and an
$f$-edge fault tolerant $2k-1$ spanner of size $O(f\cdot n^{1+1/k})$
for a graph of size $n$.
A randomized construction attaining an improved tradeoff for vertex
fault-tolerant spanners was shortly afterwards presented in \cite{DK11},
yielding (with high probability) for every graph $G = (V,E)$,
odd integer $s$ and integer $f$, an $f$-vertex fault-tolerant $s$-spanner
with $O\left(f^{2-\frac{2}{s+1}}n^{1+\frac{2}{s+1}}\log{n}\right)$ edges.
This should be contrasted with the best stretch-size tradeoff currently
known for non-fault-tolerant spanners \cite{TZ01}, namely,
$2k-1$ stretch with ${\tilde O}(n^{1+1/k})$ edges.

An efficient algorithm that given a set $V$ of $n$ points
in $d$-dimensional Euclidean space constructs an $f$-vertex
fault tolerant geometric $(1+\epsilon)$-spanner for $V$, namely,
a sparse graph $H$ satisfying that
$\dist(u,v,H\setminus F) \leq (1+\epsilon) \dist(u,v,G)$ for any set
$F\subseteq V$ of size $f$ and any pair of points $u,v \in V\setminus F$,
was presented in \cite{LNS98}. A fault tolerant geometric spanner
of improved size was later presented in \cite{L99}; finally,
a fault tolerant geometric spanner of optimal maximum degree
and total weight was presented in \cite{CZ03}.
The distinction between the stronger type of fault-tolerance obtained
for geometric graphs (termed {\em rigid} fault-tolerance)
and the more flexible type required for handling general graphs
(termed {\em competitive} fault-tolerance) is elaborated upon in \cite{P09}.

A related network service is the {\em distance oracle} \cite{BS06,RTZ05,TZ05},
which is a succinct data structure capable of supporting
efficient responses to distance queries on a weighted graph $G$.
A distance query $(s,t)$ requires finding,
for a given pair of vertices $s$ and $t$ in $V$, the distance
(namely, the length of the shortest path) between $u$ and $v$ in $G$.
The query protocol of an oracle $\cS$ correctly answers distance queries on $G$.
In a {\em fault tolerant distance oracle}, the query
may include also a set $F$ of failed edges or vertices (or both),
and the oracle $\cS$ must return, in response to a query $(s,t,F)$,
the distance between $s$ and $t$ in $G'=G\setminus F$.
Such a structure is sometimes called an {\em $F$-sensitivity distance oracle}.
The focus is on both fast preprocessing time, fast query time and low space.
It has been shown in~\cite{DTCR08} that given a directed weighted graph $G$
of size $n$, it is possible to construct in time $\Ot(mn^2)$
a $1$-sensitivity fault tolerant distance oracle of size $O(n^2 \log n)$
capable of answering distance queries in $O(1)$ time in the presence
of a single failed edge or vertex.
The preprocessing time was recently improved to $\Ot(mn)$,
with unchanged size and query time ~\cite{BK09}.
A $2$-sensitivity fault tolerant distance oracle of size $O(n^2 \log^3 n)$,
capable of answering $2$-sensitivity queries in $O(\log n)$ time,
was presented in~\cite{DP09}.
\par
Recently, distance sensitivity oracles have been considered for weighted
and directed graphs in the \emph{single source} setting \cite{GW12}.
Specifically, Grandoni and Williams considered the problem of
\emph{single-source replacement paths} where one aims to compute
the collection of all replacement paths for a given source node $s$,
and proposed an efficient randomized algorithm that does so in
$\widetilde{O}(APSP(n,M))$ where $APSP(n,M)$ is the time required to compute
all-pairs-shortest-paths in a weighted graph with integer weights $[-M,M]$.
Interestingly, although their algorithm does not aim explicitly at minimizing
the total number of edges used by the resulting collection of
replacement paths, one can show that the resulting construction yields
a rather sparse path collection, with at most $O(n^{3/2}\log n)$ edges
(although it may also be far from optimal in some instances).


Label-based fault-tolerant distance oracles for graphs of bounded clique-width
are presented in \cite{CT07}. The structure is composed of a label $L(v)$
assigned to each vertex $v$, and handles queries of the form $(L(s),L(t),F)$
for a set of failures $F$. For an $n$-vertex graph of tree-width or
clique-width $k$, the constructed labels are of size $O(k^2 \log^2 n)$.

A relaxed variant of distance oracles, in which distance queries are answered
by {\em approximate} distance estimates instead of {\em exact} ones,
was introduced in~\cite{TZ05}, where it was
shown how to construct, for a given weighted undirected $n$-vertex graph $G$,
an approximate distance oracle of size $O(n^{1+1/k})$
capable of answering distance queries in $O(k)$ time,
where the {\em stretch} (multiplicative approximation factor)
of the returned distances is at most $2k-1$.
\par
An $f$-sensitivity approximate distance oracle $\cS$ was presented in
\cite{CLPR09-do}.
For an integer parameter $k\ge 1$, the size of $\cS$ is
$O(kn^{1+\frac{8(f+1)}{k+2(f+1)}}\log{(nW)})$, where $W$ is the weight of the
heaviest edge in $G$, the stretch of the returned distance is $2k-1$,
and the query time is $O(|F|\cdot\log^2n\cdot\log\log n\cdot\log\log d)$,
where $d$ is the distance between $s$ and $t$ in $G\setminus F$.
\par
A fault-tolerant label-based $(1+\epsilon)$-approximate distance oracle
for the family of graphs with doubling dimension bounded by $\alpha$
is presented in \cite{ACGP10}.
For an $n$-vertex graph $G(V,E)$ in this family,
and for desired precision parameter $\epsilon>0$,
the distance oracle constructs and stores
an $O(\log n / \epsilon^{2\alpha})$-bit label at each vertex.
Given the labels of two end-vertices $s$ and $t$ and of collections $F_V$
and $F_E$ of failed (or ``forbidden'') vertices and edges,
the oracle computes, in time polynomial in the length of the labels,
an estimate for the distance between $s$ and $t$ in the surviving graph
$G(V\setminus F_V, E\setminus F_E)$, which approximates the true distance
by a factor of $1+\epsilon$.

Our final example concerns fault tolerant routing schemes.
A fault-tolerant routing protocol is a distributed algorithm
that, for any set of failed edges $F$, enables any source vertex $\hat s$
to route a message to any destination vertex $\hat d$ along a shortest
or near-shortest path in the surviving network $G\setminus F$
in an efficient manner (and without knowing $F$ in advance).
\par
In addition to route efficiency, it is often desirable to optimize also
the amount of memory stored in the routing tables of the vertices, possibly
at the cost of lower route efficiency, giving rise to the problem
of designing compact routing schemes
(cf. \cite{ABLP-89:stoc,S12,Peleg00:book,PU-89:tables,TZ01}).

Label-based fault-tolerant routing schemes for graphs of
bounded clique-width are presented in \cite{CT07}.
To route from $s$ to $t$, the source needs to specify the labels
$L(s)$ and $L(t)$ and the set of failures $F$, and the scheme
efficiently calculates the shortest path between $s$ and $t$ that avoids $F$.
For an $n$-vertex graph of tree-width or clique-width $k$, the
constructed labels are of size $O(k^2 \log^2 n)$.
\par
Fault-tolerant compact routing schemes are considered
in \cite{CLPR09-do}, for up to two edge failures.
Given a message $M$ destined to $t$ at a source vertex $s$, in the
presence of a failed edge set $F$ of size $|F|\le 2$ (unknown to $s$),
the scheme presented therein routes $M$ from $s$ to $t$
in a distributed manner, over a path of length at most $O(k)$ times
the length of the optimal path (avoiding $F$).
The total amount of information stored in vertices of $G$
on average is bounded by $O(k n^{1+1/k})$.
This should be compared with the best memory-stretch tradeoff currently
known for non-fault-tolerant compact routing \cite{TZ01}, namely,
$2k-1$ stretch with ${\tilde O}(n^{1+1/k})$ memory per vertex.
\par
A compact routing scheme capable of handling multiple edge failures
is presented in \cite{S12}.
The scheme routes messages (provided their source $s$ and destination $t$
are still connected in the surviving graph $G \setminus F$)
over a path whose length is proportional to the distance between $s$ and $t$
in $G \setminus F$, to $|F|^3$ and to some poly-log factor.
The routing table required at a node $v$ is of size proportional to
$v$'s degree and some poly-log factor.
\par
A routing scheme with stretch $1+\epsilon$ for graphs of bounded
doubling dimension is also presented in \cite{ACGP10}.
The scheme can be generalized also to the family of weighted graphs
of bounded doubling dimension and bounded degree.
In this case, the label size will also depend linearly on the maximum
vertex degree $\Delta$, and this is shown to be necessary.

\section{Preliminaries}
\label{sec:prelim}
\paragraph{Notation}
Given a graph $G=(V,E)$ and a source node $s$, let $T_0(s) \subseteq G$ be
a shortest paths (or BFS) tree rooted at $s$. For a source node set $S \subseteq V$,
let $T_0(S)=\bigcup_{s \in S} T_0(s)$ be a union of the single source BFS trees.
Let $\pi(s, v,T)$ be the $s-v$ shortest-path in tree $T$, when the tree $T=T_0(s)$, we may omit it and simply write $\pi(s,v)$.
Let $\Gamma(v, G)$ be the set of $v$ neighbors in $G$. Let $E(v,G)=\{(u,v) \in E(G)\}$ be the set of edges incident to $v$
in the graph $G$ and let $\deg(v,G)=|E(v,G)|$ denote the degree of node $v$ in $G$. When the graph $G$ is clear from the context,
we may omit it and simply write $\deg(v)$.
Let $\depth(s, v) = \dist(s,v,G)$ denote the {\em depth} of $v$
in the BFS tree $T_0(s)$. When the source node $s$ is clear from the context,
we may omit it and simply write $\depth(v)$.
Let $\Depth(s)=\max_{u \in V} \{ \depth(s, u) \}$ be the {\em depth}
of $T_0(s)$.
For a subgraph $G'=(V', E') \subseteq G$
(where $V' \subseteq V$ and $E' \subseteq E$)
and a pair of nodes $u,v \in V$, let $\dist(u,v, G')$ denote the
shortest-path distance in edges between $u$ and $v$ in $G'$.
For a path $P=[v_1, \ldots, v_k]$, let $\LastE(P)$ be the last edge of path $P$. Let $|P|$ denote the length of the path and $P[v_i, v_j]$ be the subpath of $P$ from $v_i$ to $v_j$. For paths $P_1$ and $P_2$, $P_1 \circ P_2$ denote the path obtained by concatenating $P_2$ to $P_1$. Assuming an edge weight function $W: E(G)\to \REAL^{+}$, let $SP(s, v_i, G, W)$ be the set of $s-v_i$ shortest-paths in $G$ according to the edge weights of $W$.  Throughout, the edges of these paths are considered to be directed away from the source node $s$. Given an $s-v$ path $P$ and an edge $e=(x,y) \in P$, let $\dist(s, e, P)$ be the distance (in edges) between $s$ and $e$ on $P$. In addition, for an edge $e=(x,y)\in T_0(s)$, define $\dist(s,e)=i$ if $\depth(x)=i-1$ and $\depth(y)=i$.
\begin{definition}
A graph $T^{*}$ is an edge (resp., vertex) \FTBFS\ tree for $G$ with respect to a source node $s \in V$, iff for every edge $f \in E(G)$ (resp., vertex $f \in V$) and for every $v \in V$,
$\dist(s, v, T^{*} \setminus \{f\})=\dist(s, v, G \setminus \{f\}).$
\par
A graph $T^{*}$ is an edge (resp., vertex) \FTMBFS\ tree for $G$ with respect to source set $S \subseteq V$, iff for every edge $f \in E(G)$ (resp., vertex $f \in V$) and for every $s \in S$ and $v \in V$,
$\dist(s, v, T^{*} \setminus \{f\})=\dist(s, v, G \setminus \{f\}).$
\end{definition}
To avoid cumbersome notation, we refer to
edge \FTBFS\ (resp., edge \FTMBFS) trees simply by \FTBFS\ (resp., \FTMBFS) trees. Throughout, we focus on edge fault, yet the entire analysis extends trivially to the case of vertex fault as well.

\paragraph{The Minimum \FTBFS\ problem}
Denote the set of solutions for the instance $(G,s)$ by
$\mathcal{T}(s,G)=\{\widehat{T} \subseteq G \mid \widehat{T} \mbox{~~is an \FTBFS\ tree w.r.t.~} s\}$.
Let $\Cost^*(s, G)=\min\{|E(\widehat{T})| ~\mid~\widehat{T} \in \mathcal{T}(s,G)\}$
be the minimum number of edges in any \FTBFS\ subgraph of $G$.
These definitions naturally extend to the multi-source case where we are
given a source set $S \subseteq V$ of size $\NSource$.
Then $$\mathcal{T}(S,G)=\{\widehat{T} \subseteq G \mid \widehat{T} \mbox{~~is a
\FTMBFS\ with respect to~} S\}$$
and $$\Cost^*(S, G)=\min\{|E(\widehat{T})| ~\mid~\widehat{T} \in \mathcal{T}(S,G)\}.$$
\par In the \emph{Minimum \FTBFS} problem we are given a graph $G$ and a
source node $s$ and the goal is to compute an \FTBFS\ $\widehat{T}\in \mathcal{T}(s,G)$
of minimum size, i.e., such that $|E(\widehat{T})|=\Cost^*(s, G)$.
Similarly, in the \emph{Minimum \FTMBFS} problem we are given a graph $G$ and a
source node set $S$ and the goal is to compute an \FTMBFS\
$\widehat{T}\in \mathcal{T}(S,G)$
of minimum size i.e., such that $|E(\widehat{T})|=\Cost^*(S, G)$.

\section{Lower Bounds}
\label{sc:lb}
In this section we establish lower bounds on the size of the \FTBFS\ and
\FTMBFS\ structures.
In Subsec. \ref{sc:lb_single} we consider the single source case and
in Subsec. \ref{sec:multiedge_lb} we consider the case of multiple sources.

\subsection{Single Source}
\label{sc:lb_single}
We begin with a lower bound for the case of a single source.
\begin{theorem}
\label{thm:lowerbound_edgeonef}
There exists an $n$-vertex graph $G(V, E)$ and a source node $s \in V$ such
that any \FTBFS\ tree rooted at $s$ has $\Omega(n^{3/2})$ edges,
i.e., $\Cost^*(s,G)=\Omega(n^{3/2})$.
\end{theorem}
\Proof
Let us first describe the structure of the graph $G=(V,E)$.
Set $d=\lfloor \sqrt{n}/2 \rfloor$.

The graph consists of four main components.
The first is a path $\pi=[s=v_1, \ldots, v_{d+1}=v^*]$ of length $d$.
The second component consists of a node set $Z=\{z_1,\ldots,z_d\}$
and a collection of $d$ disjoint paths of deceasing length, $P_1, \ldots, P_d$,
where $P_j=[v_j=p^j_1, \ldots, z_j=p^j_{t_j}]$ connects $v_j$ with $z_j$ and its length is $t_{j}=|P_j|=6+2(d-j)$, for every $j \in 1,\cdots,d$.
Altogether, the set of nodes in these paths, $Q=\bigcup_{j=1}^d V(P_j)$,
is of size $|Q| = d^2+7d$.
The third component is a set of nodes $X$ of size $n-(d^2+7d)$,
all connected to the terminal node $v^*$.
The last component is a complete bipartite graph $B=(X, Z, \hat{E})$
connecting $X$ to $Z$.
Overall, $V=X \cup Q$ and $E= \hat{E} \cup E(\pi) \cup \bigcup_{j=1}^d E(P_j)$.
Note that $n/4 \le |Q| \le n/2$ for sufficiently large $n$. Consequently,
$|X| = n- |Q| \ge n/2$, and $|\hat{E}| = |Q|\cdot |X| \ge n^{3/2}/4$.

\begin{figure}[htb!]
\begin{center}
\includegraphics[scale=0.36]{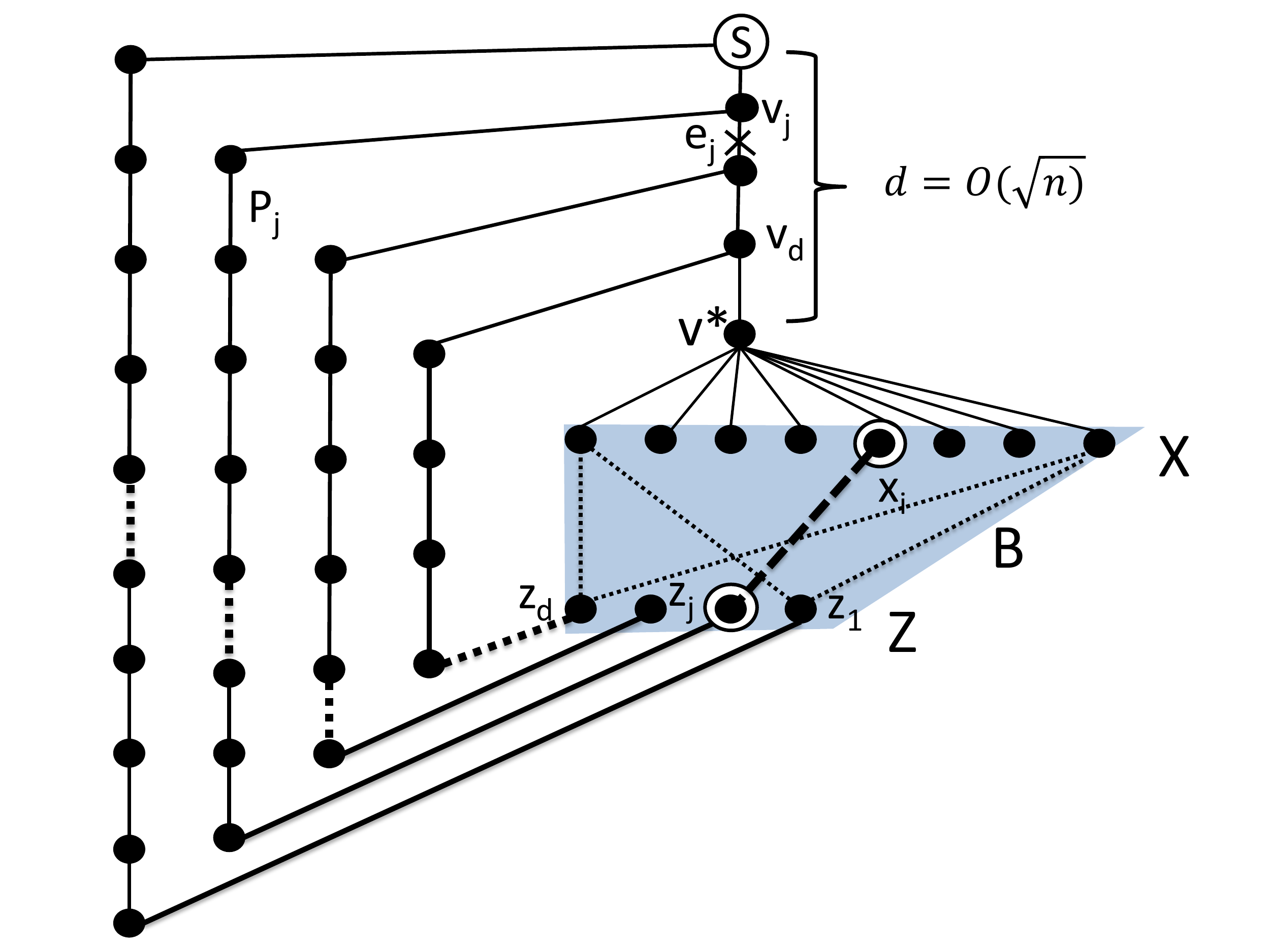}
\caption{Lower bound construction for \FTBFS\.
The original BFS tree consists of the non-dashed edges.
The dashed edges are the ones necessary to make it an \FTBFS\ tree.
For example, the bold dashed edge $(x_j, z_i)$ is required
upon failure of the edge $e_i$. \label{fig:lowerbound1f}}
\end{center}
\end{figure}

A BFS tree $T_0$ rooted at $s$ for this $G$
(illustrated by the solid edges in Fig. \ref{fig:lowerbound1f})
is given by
$$E(T_0) ~=~ \{(x_i, z_i) \mid i \in \{1,\ldots, d\}\} \cup
\bigcup_{j=1}^d E(P_j) \setminus \{(p^j_{\ell_j}, p^j_{\ell_j-1})\},$$
where $\ell_j=t_j-(d-j)$ for every $j \in \{1, \ldots, d\}$.
We now show that every \FTBFS\ tree $T' \in \mathcal{T}(s,G)$ must contain
all the edges of $B$, namely, the edges $e_{i,j}=(x_i, z_j)$
for every $i \in \{1, \ldots, |X| \}$ and $j \in \{1, \ldots, d\}$
(the dashed edges in Figure \ref{fig:lowerbound1f}).
Assume, towards contradiction, that there exists
a $T' \in \mathcal{T}(s,G)$ that does not contain $e_{i,j}$
(the bold dashed edge $(x_i, z_j)$ in the figure).
(the bold dashed edge $(x_i, z_j)$ in Figure \ref{fig:lowerbound1f}).
Note that upon the failure of the edge $e_j=(v_j,v_{j+1}) \in \pi$,
the unique $s-x_i$ shortest-path connecting $s$ and $x_i$ in
$G \setminus \{e_j\}$ is $P'_{j}= \pi[v_1,v_j] \circ P_{j}\circ [z_{j}, x_i]$,
and all other alternatives are strictly longer.
%
Since $e_{i,j} \notin T'$, also $P'_j \nsubseteq T'$, and therefore
$\dist(s, x_i, G \setminus \{e_j\})< \dist(s, x_i, T' \setminus \{e_j\})$,
in contradiction to the fact that $T'$ is an \FTBFS\ tree.
It follows that every \FTBFS\ tree $T'$ must contain at least
$|\hat{E}| = \Omega(n^{3/2})$ edges. The theorem follows.
\QED

\subsection{Multiple Sources}
\label{sec:multiedge_lb}
\begin{figure}[h!]
\begin{center}
\includegraphics[scale=0.4]{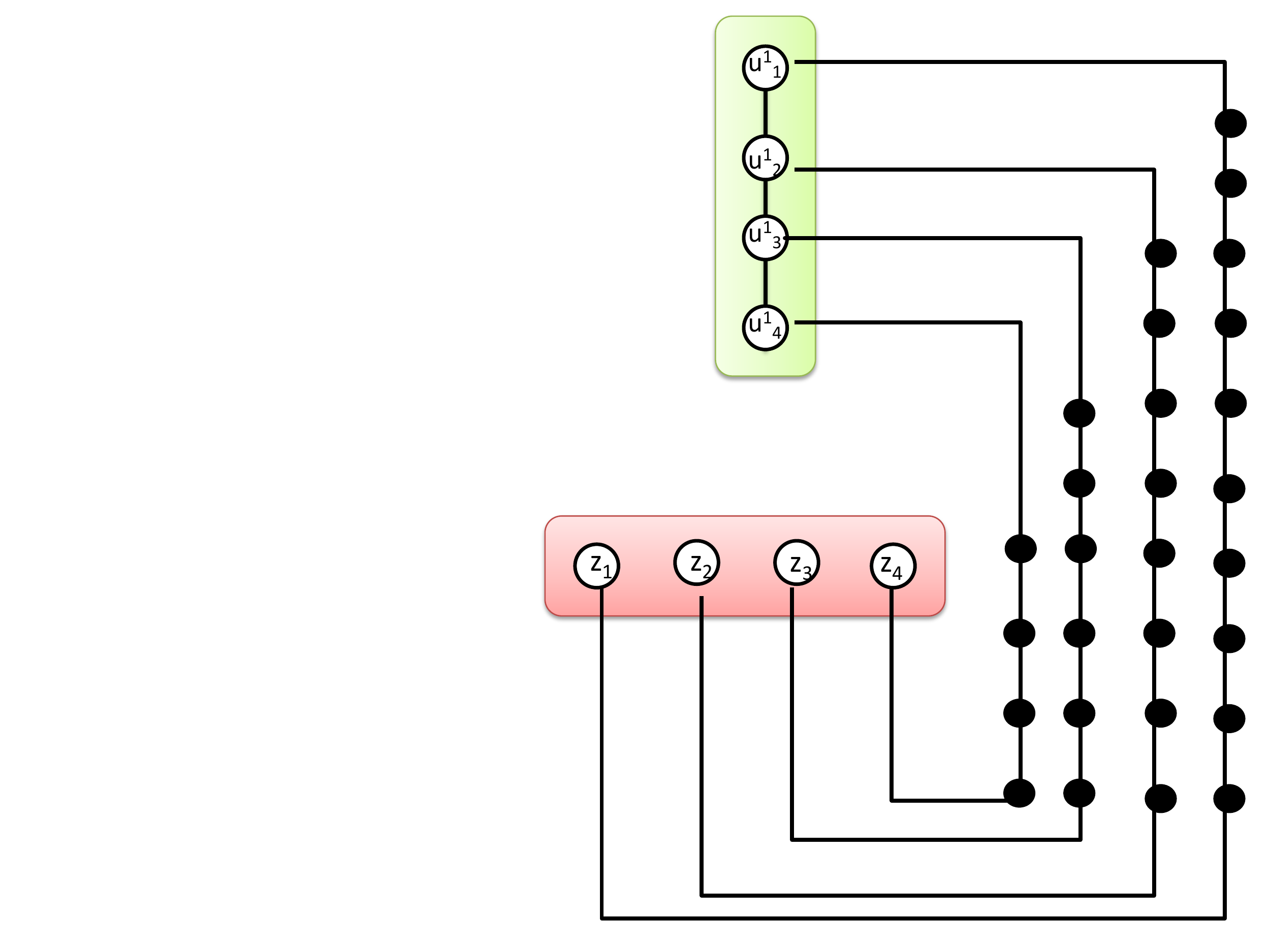}
\caption{\label{fig:lowerboundg1} The graph $G_1(d)$.}
\end{center}
\end{figure}

We next consider an intermediate setting where it is necessary to construct
a fault-tolerant subgraph \FTMBFS\ containing several \FTBFS\ trees in parallel,
one for each source  $s \in S$, for some $S\subseteq V$.
We establish the following.
\begin{theorem}
\label{thm:lowerbound_f_multisource}
There exists an $n$-vertex graph $G(V, E)$ and a source set $S \subseteq V$
of cardinality $\NSource$, such that any \FTMBFS\ tree from
the source set $S$ has $\Omega(\sqrt{\NSource}\cdot n^{3/2})$ edges,
i.e., $\Cost^*(S,G)=\Omega(\sqrt{\NSource}\cdot n^{3/2})$.
\end{theorem}
%
%
\Proof
Our construction is based on the graph $G(d)=(V_1,E_1)$,
which consists of three components:
(1) a set of vertices $U=\{u_1,\ldots,u_d\}$ connected by a path
$P_1=[u_1, \ldots, u_d]$,
(2) a set of terminal vertices $Z=\{z_1,\ldots,z_d\}$
(viewed by convention as ordered from left to right),
and
(3) a collection of $d$ vertex disjoint paths $Q_{i}$ of length
$|Q_i|=6+2 (d-i)$ connecting $u_i$ and $z_i$
for every $i \in \{1, \ldots, d\}$.
Thus $|Q^1_1|> \ldots> |Q^1_d|$.
The vertex $\Root(G(d))=u_d$ is fixed as the root of $G(d)$, hence
the edges of the paths $Q_i$ are viewed as directed away from $u_i$,
and the terminal vertices of $Z$ are viewed as the \emph{leaves} of the graph,
denoted $\Leaf(G(d))=Z$.
See Fig. \ref{fig:lowerboundg1} for illustration.

Overall, the vertex and edge sets of $G(d)$ are
$V_1=U \cup Z \cup \bigcup_{i=1}^d V(Q_i)$ and
$E_1=E(P_1) \cup \bigcup_{i=1}^d E(Q_i)$.

\begin{observation}
\label{obs:rel}
\begin{description}
\item{(a)}
The number of leaves in $G(d)$ is $|\Leaf(G(d))|=d$.
\item{(b)}
$|V_1|=c \cdot d^{2}$ for some constant $c$.
\end{description}
\end{observation}

\begin{figure}[h!]
\begin{center}
\includegraphics[scale=0.5]{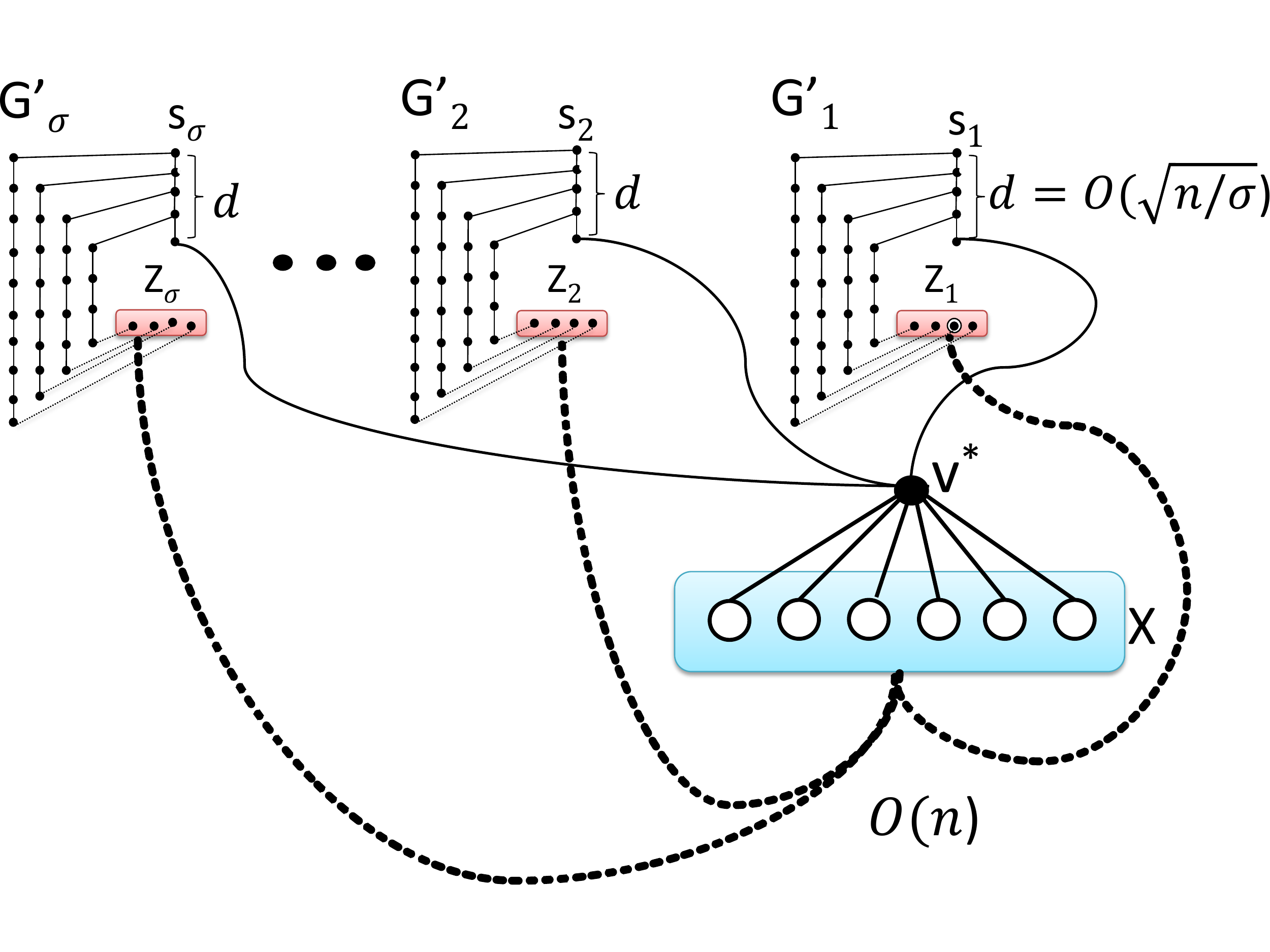}
\caption{\label{fig:lowerboundmulti}
Illustration of the lower bound for the multi-source case.}
\end{center}
\end{figure}

Take $\NSource$ copies, $G'_1, \ldots, G'_\NSource$, of $G(d)$,
where $d=O((n / \NSource)^{1/2})$.
Note that Obs. \ref{obs:rel}, each copy $G'_i$ consists of $O(n/\NSource)$
nodes.
Let $y_i$ be the node $u_d$ and $s_i=\Root(G'_i)$ in the $i$th copy $G'_i$.
Add a node $v^*$ connected to a set $X$ of $\Omega(n)$ nodes and connect $v^*$
to each of the nodes $y_i$, for $i \in \{1, \ldots, d\}$.
Finally, connect the set $X$ to the $\NSource$ leaf sets
$\Leaf(G'_1), \ldots, \Leaf(G'_\NSource)$ by a complete bipartite graph,
adjusting the size of the set $X$ in the construction so that $|V(G)|=n$.
Since $\NLeaf(G'_i)=\Omega((n / \NSource)^{1/2})$ (see Obs. \ref{obs:rel}),
overall $|E(G)| = \Omega(n \cdot \NSource \cdot \NLeaf(G_1(d))) =
\Omega(n\cdot(\NSource n)^{1/2})$.
Since the path from each source $s_i$ to $X$ cannot aid the nodes of $G'_j$
for $j \neq i$, the analysis of the single-source case can be applied
to show that each of the bipartite graph edges in necessary
upon a certain edge fault.
See Fig. \ref{fig:lowerboundmulti} for an illustration.
\QED

\section{Upper Bounds}
\label{sec:upper}
In this section we provide tight matching upper bounds to the lower bounds
presented in Sec. \ref{sc:lb}.

\subsection{Single Source}
\label{subsec:1edge_alg}
For the case of \FTBFS\ trees, we establish the following.
\begin{theorem}
There exists an $O(n m)$ time algorithm that for every $n$-vertex graph $G$
and source node $s$ constructs an \FTBFS\ tree rooted at $s$ with
$O(n \cdot \min\{\Depth(s), \sqrt{n}\})$ edges.
\end{theorem}
To prove the theorem, we first describe a simple algorithm for the problem
and then prove its correctness and analyze the size
of the resulting \FTBFS\ tree.
We note that using the sparsity lemma of \cite{RTREP05} and the tools of
\cite{GW12}, one can provide a randomized construction for an \FTBFS\ tree
with $O(n^{3/2} \log n)$ edges with high probability.
In contrast, the algorithm presented in this paper is
\emph{deterministic} and achieve an \FTBFS\ tree with $O(n^{3/2})$ edges,
matching exactly the lower bound established in Sec. \ref{sc:lb}.

\paragraph{The Algorithm}
To avoid complications due to shortest-paths of the same length,
we assume all shortest-path are computed with a weight assignment $W$
that guarantees the uniqueness of the shortest-paths. This can be achieved
by considering
a weight function $W$ defined so as to ensure that the shortest paths are also
of minimal number of edges but at the same time guarantees the uniqueness
of the $u-v$ shortest-path, for every $u,v \in V$. Let $e_1, \ldots, e_{m}$ be some arbitrary ordering of $E(G)$. Then set $W(e_k) = 2^{m+1} + 2^k$.
Let $T_0=BFS(s, G)$ be the BFS tree rooted at $s$ in $G$, computed according
to the weight assignment $W$. For every $e_j \in T_0$, let $T_0(e_j)$ be the
BFS tree rooted at $s$ in $G \setminus \{e_{j}\}$.
Then the final \FTBFS\ tree is given by
$$T^{*}(s)=T_0\cup \bigcup_{e_j \in T_0} T_0(e_j).$$
The correctness is immediate by construction.
\begin{observation}
\label{obs:correctness}
$T^{*}(s)$ is an \FTBFS\ tree.
\end{observation}
\def\APPENDUPCORRECT{
\Proof
Consider a vertex $v$ and an edge $e$.
If $e \notin \pi(s, v)$,
then $\pi(s, v) \subseteq T^{*}(s)\setminus\{e\}$, hence $\dist(s, v, T^{*}(s) \setminus \{e\})=|\pi(s, v)|=\dist(s, v, G \setminus \{e\})$. Otherwise, $e \in \pi(s, v) \subseteq T_0$. Then by construction, $T_0(e) \subseteq T^{*}(s)$. By definition, $\dist(s, v, T^*(s) \setminus \{e\})=\dist(s, v, T_0(e))=\dist(s, v, G \setminus \{e\})$. The observation follows.
\QED
}
\APPENDUPCORRECT

Due to \cite{TH99} each of the $n-1$ BFS trees $T_0(e_j)$ can be constructed in $O(m)$ time, hence $O(n m)$ rounds are required in total.
It therefore remains to bound the size of $T^{*}(s)$.

\paragraph{Size Analysis}
We first provide some notation.
For a path $P$, let $\Cost(P)=\sum_{e \in P} W(e)$ be the weighted cost of $P$, i.e., the sum of its edge weights.
An edge $e \in G$ is defined as \emph{new} if $e \notin E(T_0)$.
For every $v_i \in V$ and $e_j \in T_0$, let
$P^*_{i,j}=\pi(s, v_i, T_0(e_j)) \in SP(s, v_i, G \setminus \{e_j\}, W)$
be the optimal \emph{replacement path} of $s$ and $v_i$ upon
the failure of $e_j \in T_0$.
Let $\New(P)=E(P) \setminus E(T_0)$ and
$$\New(v_i) ~=~ \{\LastE(P^*_{i,j}) \mid e_j \in T_0\} \setminus E(T_0)$$
be the set of $v_i$ new edges appearing as the last edge in the replacement paths $P^{*}_{i,j}$ of $v_i$ and $e_j \in T_0$.
It is convenient to view the edges of $T_0(e_j)$ as directed away from $s$.
We then have that
$$T^{*}=T_0\cup \bigcup_{v_i \in V \setminus \{s\}} \New(v_i).$$
I.e., the set of new edges that participate in the final \FTBFS\ tree $T^{*}$
are those that appear as a last edge in some replacement path.
\par We now upper bound the size of the \FTBFS\ tree $T^{*}$.
Our goal is to prove that $\New(v_i)$ contains at most $O(\sqrt{n})$ edges for every $v_i \in V$.
The following observation is crucial in this context.
\begin{observation}
\label{obs:new_edge}
If $\LastE(P^*_{i,j})\notin E(T_0)$, then
$e_j \in \pi(s,v_i)$.
\end{observation}
\def\APPENDONENEWEDGE{
\Proof
Assume, towards contradiction, that $e_j \notin \pi(s,v_i)$ and let
$P^*_{i,j} \subseteq T_0(e_j)$ be the $s-v_i$ replacement path in
$G \setminus \{e_j\}$ according to the weight assignment $W$.
Since $\LastE(P^*_{i,j}) \notin E(T_0)$,
we have two different $s-v_i$ shortest paths in $G \setminus \{e_j\}$, namely,
$\pi(s,v_i)$ and
$P^*_{i,j}$. By the optimality of $\pi(s,v_i)$ in $G$, i.e.,
$\pi(s,v_i) \in SP(s, v_i, G, W)$,
it holds that $\Cost(\pi(s,u)) < \Cost(P^*_{i,j})$.
On the other hand, by the optimality of $P^*_{i,j}$ in $G \setminus \{e_j\}$, i.e.,
$P^*_{i,j} \in SP(s, v_i, G \setminus \{e_j\}, W)$, we have that
$\Cost(\pi(s,u))> \Cost(P^*_{i,j})$. Contradiction.
\QED
}
\APPENDONENEWEDGE

Obs. \ref{obs:new_edge} also yields the following.
\begin{corollary}
(1) $\New(v_i) ~=~\{\LastE(P^*_{i,j})~\mid~ e_j \in \pi(s,v_i)\} \setminus E(T_0)$
and \\
(2) $|\New(v_i)| \leq \min\{\depth(v_i), \deg(v_i)\}.$
\end{corollary}
This holds since the edges of $\New(v_i)$ are coming from at most
$\depth(v_i)$ replacement paths $P^*_{i,j}$
(one for every $e_j \in \pi(s, v_i)$),
and each such path contributes at most one edge incident to $v_i$.
\par For the reminder of the analysis, let us focus on
one specific node $u=v_i$ and let  $\pi=\pi(s,u)$, $N=|\New(u)|$.
For every edge $e_k \in \New(u)$, we define the following parameters.
Let $f(e_k) \in \pi$ be the failed edge such that $e_k \in T_0(f(e_k))$
appears in the replacement path $P_k=\pi(s, u,T')$ for $T'= T_0(f(e_k))$.
(Note that $e_k$ might appear as the last edge on the path $\pi(s, u, T_0(e'))$
for several edges $e' \in \pi$; in this case, one such $e'$ is chosen arbitrarily).

Let $b_k$ be the \emph{last} divergence point of
$P_k$ and $\pi$, i.e., the last vertex on the
replacement path $P_k$ that belongs to
$V(\pi)\setminus \{u\}$.
Since $\LastE(P_k) \notin E(T_0)$,
it holds that $b_k$ is not the neighbor of $u$ in $P_k$.
\par Let $\New(u)=\{e_1, \ldots, e_N\}$ be sorted in non-decreasing order
of the distance between $b_k$ and $u$, $\dist(b_{k}, u,\pi)=|\pi(b_k, u)|$.
I.e.,
\begin{equation}
\label{eq:order}
\dist(b_1, u,\pi)\leq \dist(b_2, u,\pi) \ldots \leq \dist(b_N, u,\pi).
\end{equation}
We consider the set of truncated paths $P'_k=P_k[b_k, u]$ and show that these paths are vertex-disjoint except for the last common endpoint $u$.
We then use this fact to bound the number of these paths, hence bound the number $N$ of new edges. The following observation follows immediately by the definition of $b_k$.
\begin{observation}
\label{obs:notinpath}
$(V(P'_k)\cap V(\pi)) \setminus \{b_k,u\}=\emptyset$.
\end{observation}
\begin{lemma}
\label{cl:disjointpathmulti}
$\left(V(P'_i)\cap V(P'_j) \right) \setminus \{u\}=\emptyset$ for every
$i,j \in \{1, \ldots, N\}$, $i \neq j$.
\end{lemma}
\Proof
Assume towards contradiction that there exist $i \neq j$, and a node
$$u' \in \left(V(P'_i)\cap V(P'_j) \right) \setminus \{u\}$$ in the intersection.
Since $\LastE(P'_i)\neq \LastE(P'_j)$, by Obs. \ref{obs:notinpath} we have that $P'_i,P'_j \subseteq G \setminus E(\pi)$.
The faulty edges $f(e_i), f(e_j)$ belong to $E(\pi)$.
Hence there are two distinct $u'-u$ shortest paths in $G \setminus \{f(e_i), f(e_j)\}$.
By the optimality of $P'_i$ in $T_0(f(e_i))$,
(i.e., $P_i \in SP(s, u, G \setminus \{f(e_i)\},W)$), we have that
$\Cost(P'_i[u',u])<\Cost(P'_j[u',u])$.
In addition, by the optimality of
$P'_j$ in $T_0(f(e_j))$, (i.e., $P_j \in SP(s, u, G \setminus \{f(e_j)\},W)$),
we have that $\Cost(P'_j[u',u])<\Cost(P'_i[u',u])$.
Contradiction.
\QED
We are now ready to prove our key lemma.
\begin{lemma}
\label{lem:upper_bound}
$|\New(u)|=O(n^{1/2})$ for every $u \in V$.
\end{lemma}
\Proof
Assume towards contradiction that $N=|\New(u)|>\sqrt{2n}$.
By Lemma \ref{cl:disjointpathmulti}, we have that $b_1, \ldots, b_N$ are distinct and by definition they all appear on the path $\pi$.
Therefore, by the ordering of the $P'_k$, we have that the
inequalities of Eq. (\ref{eq:order}) are strict, i.e.,
$\dist(b_1, u,\pi)<\dist(b_2, u,\pi)< \ldots < \dist(b_N, u,\pi)$.
Since $b_1 \neq u$ (by definition),
we also have that
$\dist(b_1, u,\pi)\geq 1$. We Conclude that
\begin{equation}
\label{eq:depthprot}
\dist(b_k, u,\pi)=|\pi(b_k, u)|\geq k~.
\end{equation}
Next, note that each $P'_k$ is a replacement $b_k-u$ path and hence it cannot be shorter than $\pi(b_k, u)$, implying that $|P'_k|\geq |\pi(b_k, u)|$. Combining, with Eq. (\ref{eq:depthprot}), we have that
\begin{equation}
\label{eq:depthprotk}
|P'_k| \geq k \mbox{~~for every~~} k \in \{1, \ldots, N\}~.
\end{equation}
Since by Lemma \ref{cl:disjointpathmulti}, the paths $P'_k$ are vertex disjoint (except for the common vertex $u$), we have that
\begin{eqnarray*}
\left |\bigcup_{k=1}^N (V(P'_k)\setminus \{u\}) \right|&=&\sum_{k=1}^N |V(P'_k)\setminus \{u\}|~\geq~ \sum_{k=1}^N (k-1)~ > ~n,
\end{eqnarray*}
where the first inequality follows by Eq. (\ref{eq:depthprotk}) and the last inequality by the assumption that $N>\sqrt{2n}$.
Since there are a total of $n$ nodes in $G$, we end with contradiction.
\QED

Turning to the case of a single vertex failure, the entire
proof goes through almost without change, yielding the following.
\begin{theorem}
\label{thm:vertexonef}
There exists a polynomial time algorithm that for every $n$-vertex graph
and source node $s$ constructs  an \FTBFS\ tree from $s$
tolerant to one vertex failure,
with $O(n \cdot \min\{\Depth(s), \sqrt{n}\})$ edges.
\end{theorem}

\subsection{Multiple Sources}
\label{subsec:multisource}
For the case of multiple sources, we establish the following upper bound.
\begin{theorem}
\label{thm:multi_source_edgeonef}
There exists a polynomial time algorithm that for every $n$-vertex graph
$G=(V,E)$ and source set $S \subseteq V$ of size $|S|=\NSource$ constructs
an \FTMBFS\ tree $T^*(S)$ from each source $s_i \in S$,
tolerant to one edge or vertex failure,
with a total number of
$n\cdot \min\{\sum_{s_i \in S} \depth(s_i),O(\sqrt{\NSource n})\}$ edges.
\end{theorem}
\def\APPENDUPPERMULTI{

\paragraph{The algorithm}
As in the single source case, to avoid complications due to shortest-paths
of the same length, all shortest path distances in $G$ are computed using
a weight function $W$ defined so as to ensure the uniqueness of a single $u-v$
shortest-path.
For every $s_i \in S$ and every $e_{j} \in T_0(s_i)$, let $T(s_i, e_j)$ be the BFS tree rooted at $s_i$ in $G \setminus \{e_j\}$.
Let $$T_0(S)=\bigcup_{s_i \in S}T_0(s_i)$$
be the joint structure containing all the BFS trees of $S$.
Then by the previous section, the \FTBFS\ tree for $s_i$ is $T^*(s_i)=T_0 \cup \bigcup_{e_j \in T_0(s_i)}T(s_i, e_j)$. Define the \FTMBFS\ for $S$ as
$$T^*(S)=\bigcup_{s_i \in S}T^*(s_i)=\bigcup_{s_i \in S, e_{j} \in T_0(s_i)} T(s_i, e_j).$$

\paragraph{Analysis}
The correctness follows immediately by the single source case.
It remains to bound the number of edges of $T^*(S)$.
An edge $e$ is \emph{new} if $e \notin  T_0(S)$.
For every $v_i \in V$, define its new edge set in the graph $T^*(S)$ by
$$\New(S,v_i)=\{\LastE(\pi(s, v_i, T(s_i, e_j)))~\mid~ s_i \in S, e_j \in T_0(s_i)\} \setminus E(T_0(S)).$$
To bound the size of $T^*(S)$, we focus on node $u=v_i$,
and bound its new edges $\New(S,u)=\{e_1, \ldots, e_N\}$.
Obs. \ref{obs:new_edge} yields the following.
\begin{corollary}
\label{cor:multi_depth}
$\New(S,u) \leq \sum_{s_i \in S} \depth(s_i)$.
\end{corollary}
Towards the end of this section, we prove that $\New(S,u)$ contains at most $O(\sqrt{\NSource n})$ new edges.
For ease of notation, let $\pi(s_i)=\pi(s_i, u)$ for every
$i \in \{1, \ldots, \NSource\}$. For every edge $e_k \in \New(S,u)$,
we define the following parameters. Let $s(e_k) \in S$ and
$f(e_k) \in T_0(s(e_k))$ be such that $e_k\in T(s(e_k), f(e_k))$.
I.e., the edge $e_k$ appears in the replacement $s(e_k)-u$ path
$P_k=\pi(s, u, T')$, where $T'=T(s(e_k), f(e_k))$ is
the BFS tree rooted at $s(e_k)$ in $G \setminus \{f(e_k)\}$.
By Obs. \ref{obs:new_edge}, $f(e_k) \in \pi(s(e_k))$.
(Note that for a given new edge $e_k$ there might be several $s'$
and $e'$ such that $e_k=\LastE(\pi(s', u, T(s',e')))$; in this case one such pair $s',e'$ is chosen arbitrarily.)
For every replacement path $P_k$ (whose last edge is $e_k$), denote by
$b_k$ the \emph{last} divergence point of $P_k$ and the collection of shortest
$s_i-u$ paths $\mathcal{P}=\bigcup_{s_i \in S} \pi(s_i,u) \setminus \{u\}$.
I.e., $b_k$ is the last point on $P_k$ that belongs to $V(\mathcal{P})\setminus \{u\}$.
Let $P'_k=P_k[b_k,u]$ be the truncated path from the divergence point $b_k$ to $u$.
Note that since $e=(x,u)=\LastE(P_k) \notin E(T_0(S))$ is a new edge,
it holds that $x \notin V(\mathcal{P})\setminus \{u\}$ and $b_k$ is
in $V \setminus \{u\}$.
The following observation is useful.
\begin{observation}
\label{obs:pathmulti}
$P'_k \subseteq G \setminus E(\mathcal{P})$ for every $k \in \{1, \ldots, N\}.$
\end{observation}
We now show that the paths $P'_k$ are vertex disjoint
except for their endpoint $u$ (this is regardless of their respective source
$s(e_k)$).
\begin{lemma}
\label{cl:disjointpathmulti2}
$\left(V(P'_i)\cap V(P'_j) \right) \setminus \{u\}=\emptyset$ for every $i \neq j \in \{1, \ldots, N\}$.
\end{lemma}
\Proof
Assume towards contradiction that there exists $i \neq j$,
and a node $$u' \in \left(V(P'_i)\cap V(P'_j) \right) \setminus \{u\}$$
in the intersection. Since $\LastE(P'_i)\neq \LastE(P'_j)$ and
by Obs. \ref{obs:pathmulti}, $P'_i,P'_j \subseteq G \setminus E(\mathcal{P})$,
and the faulty edges $f(e_i), f(e_j) \in \mathcal{P}$,
we have two distinct $u'-u$ replacement paths in $G \setminus \{f(e_i), f(e_j)\}$.
By the optimality of $P'_i$ in $T(s(e_i), f(e_i))$,
(i.e.,  $P_i \in SP(s(e_i), u, G\setminus \{f(e_i)\},W)$),
we have that $\Cost(P'_i)<\Cost(P'_j)$.
Similarly, by the optimality of $P'_j$ in $T(s(e_j), f(e_j))$,
(i.e.,  $P_j \in SP(s(e_j), u, G\setminus \{f(e_j)\},W)$),
we have that $\Cost(P'_j)<\Cost(P'_i)$, contradiction. The lemma follows.
\QED
We are now ready to state and prove our main lemma.
\begin{lemma}
\label{cl:multi_nnew_node}
$N=|\New(S, u)|=O(\sqrt{\NSource n})$.
\end{lemma}
We begin by classifying the set of new edges $e_i \in \New(S,u)$ into $\NSource$
classes according to the position of the divergence point $b_i$.
For every $e_i \in \New(S,u)$, let $\widehat{s}(e_i)\in S$ be some source node such that the divergence point $b_i \in \pi(\widehat{s}(e_i),u)$ appears on its $\widehat{s}(e_i)-u$ shortest path $T_0(S)$.
If there are several such sources for the edge $e_i$,
one is chosen arbitrarily.
\par For every $s_j \in S$, let
$$\New(s_j)=\{e_i \in \New(S,u) \mid \widehat{s}(e_i)=s_j\}$$
be the set of new edges in $\New(S,u)$ that are mapped to $s_j \in S$.
Then, $\New(S,u)=\bigcup_{s_j \in S}\New(s_j)$.
Let $x_j=|\New(s_j)|$.
\par We now focus on $s_j$. For every $e_{j_k} \in \New(s_j)$, $k=\{1, \ldots, x_j\}$, let $P_{j_k}=\pi(s(e_{j_k}), u, T')$ for $T'=T(s(e_{j_k}), f(e_{j_k}))$ be the replacement path such that $\LastE(P_{j_k})=e_{j_k}$ and $b_{j_k}$ be its corresponding (last) divergence point with $\pi(s_j, u)$ ($s_j=\widehat{s}(e_{j_k})$). In addition, the truncated path is given by $P'_{j_k}=P_{j_k}[b_{j_k},u]$. Note that $\LastE(P_{j_k})=e_{j_k}$.
\par Consider the set of divergence points $b_{j_1}, \ldots, b_{j_{x_j}}$ sorted in non-decreasing order of the distance between $b_{j_k}$ and $s_j$ on the shortest $s_j-u$ path $\pi(s_j)$ i.e., $|\pi(b_{j_k},u,T_0(s_j))|$, where
\begin{equation}
\label{eq:sorted_multi}
|\pi(b_{j_1}, u,T_0(s_j))| \leq |\pi(b_{j_2}, u,T_0(s_j))| \ldots \leq |\pi(b_{j_{x_j}}, u, T_0(s_j))|~.
\end{equation}
Note that by Lemma \ref{cl:disjointpathmulti2}, $b_{j_\ell} \neq b_{j_{\ell'}}$ for every $\ell, \ell' \in \{1, \ldots, x_j\}$. In addition, since each $b_{j_\ell} \neq u$, $|\pi(b_{j_1}, u,T_0(s_j))|\geq 1$.
Hence, since $b_{j_1}, \ldots, b_{j_{x_j}} \in \pi(s_j)$, combining with Eq. (\ref{eq:sorted_multi}) we get that
\begin{equation}
\label{eq:sorted_multi2}
1\leq |\pi(b_{j_1}, u,T_0(s_j))| < |\pi(b_{j_2}, u,T_0(s_j))| \ldots < |\pi(b_{j_{x_j}}, u,T_0(s_j))|~.
\end{equation}
Since $P'_{j_\ell}$ is an alternative $b_{j_\ell}-u$ replacement path, we have that
\begin{equation}
\label{eq:sorted_multi_spath}
|P'_{j_\ell}|\geq |\pi(b_{j_\ell}, u,T_0(s_j))| \geq \ell.
\end{equation}
where the last inequality follows by Eq. (\ref{eq:sorted_multi}).
Hence, since all $P'_{j_\ell}$ are vertex disjoint, except for the last node $u$, we get the total number of nodes $V(s_j)=\bigcup V(P'_{j_{\ell}}) \setminus \{u\}$
occupied by $P'_{j_\ell}$ paths is
$$\sum_{\ell=1}^{x_{j}} |V(P'_{j_\ell})|= |V(s_{j})|=O(x_j^2).$$
Since the nodes of $V(s_{j_1})$ and $V(s_{j_2})$ are disjoint for every $s_{j_1}, s_{j_2} \in S$, by Lemma \ref{cl:disjointpathmulti2},
it follows that $|\New(S, u)|=\sum_{j=1}^{\NSource} x_j$ but $\sum_{j=1}^{\NSource} |V(s_j)|=O(x_j^2) \leq n$. Therefore, $|\New(S,u)|=\sum_{j=1}^\NSource x_j \leq O(\sqrt{\NSource n})$.
\QED
As there are $n$ nodes, combining with Cor. \ref{cor:multi_depth}, we get that the total number of edges in $T^*(S)$ is given by
$$E(T^*(S)) \leq |E(T_0(S))|+ \sum_{u \in V} |\New(S,u)| \leq \NSource n + n \cdot \min\{\sum_{s_i \in S} \depth(s_i),O(\sqrt{\NSource n})\},$$
as required. Thm. \ref{thm:multi_source_edgeonef} is established.
The analysis for the case of vertex faults follows with almost no changes.
\QED
}
\APPENDUPPERMULTI
\section{Hardness of Approximation of the Minimum \FTBFS\ Problem}
\label{sec:hardness}
In this section we establish the following.
\begin{theorem}
\label{thm:hardness}
The Minimum \FTBFS\ problem is NP-complete and cannot be approximated to within a factor $c \log n$ for some constant $c>0$ unless $\mathcal{NP} \subseteq  \mathcal{TIME}(n^{poly\log(n)})$.
\end{theorem}
We prove Theorem \ref{thm:hardness} by showing a gap preserving reduction
from the Set-Cover problem to the Minimum \FTBFS\ problem.
An instance $\langle U, \Set \rangle$ of the Set-Cover problem consists of
a set of $N$ elements $U=\{u_1, \ldots, u_N\}$ and
a collection of $M$ sets $\Set=\{S_1, \ldots, S_M\}$
such that $S_i \subseteq U$ and $\bigcup S_i=U$.
The task is to choose the minimal number of sets
in $\Set$ whose union covers all of $U$.
Fiege \cite{Feige98} showed that the Set Cover problem cannot be approximated
within a ratio of $(1-o(1)) \ln n$ unless $\mathcal{NP} \subseteq  \mathcal{TIME}(M^{poly\log(M)})$.

\begin{figure}[ht!]
\begin{center}
\includegraphics[scale=0.5]{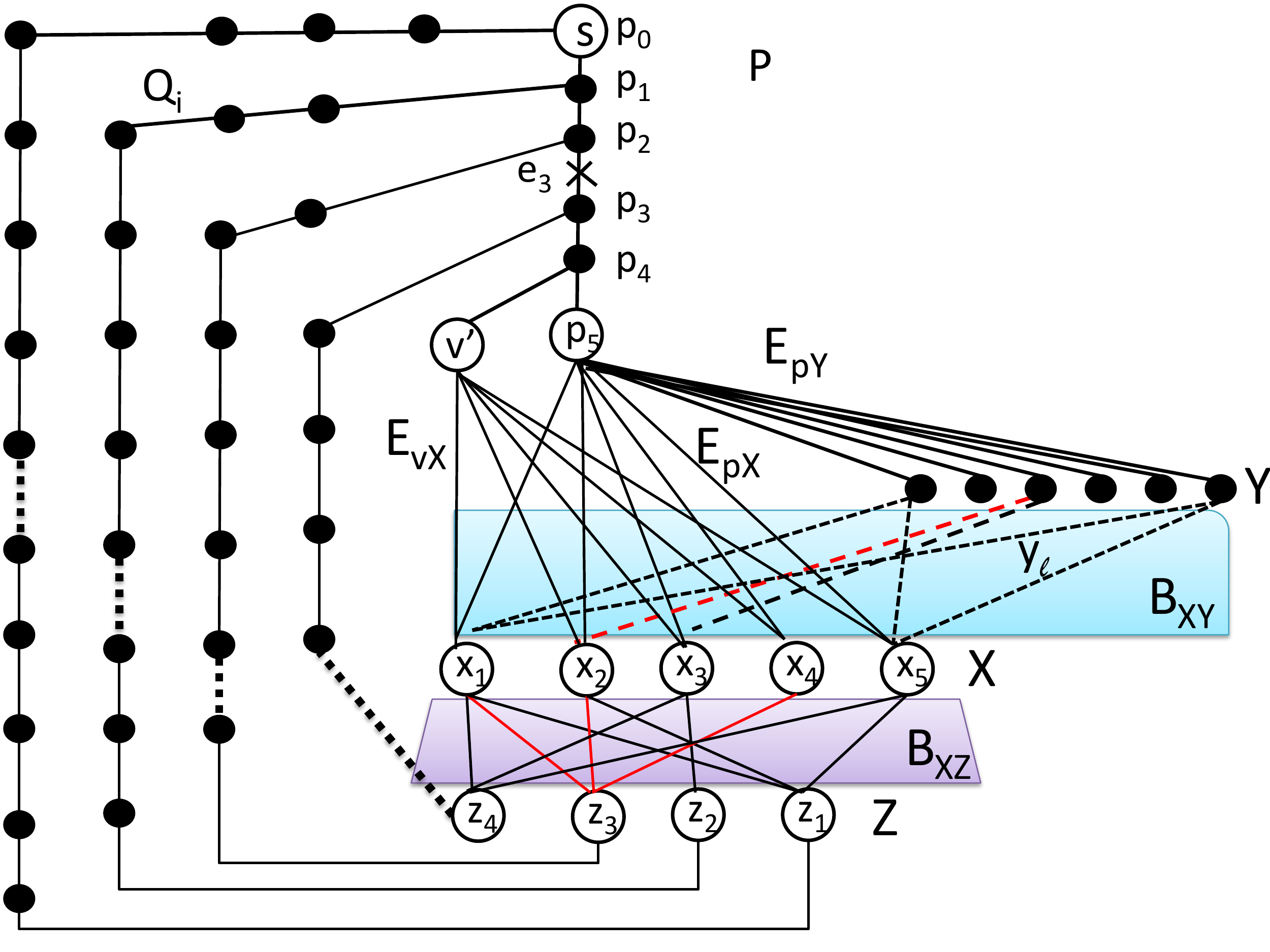}
\caption{
\label{fig:reducbfs}
Schematic illustration of the reduction
from Set-Cover to Minimum $\FTBFS$.
In this example $\Set=\{S_1, S_2, \ldots, S_5\}$
where $S_1=\{u_1, u_3, u_4\}$, $S_2=\{u_1, u_3\}$, $S_3=\{a_2, a_4\}$,
$S_4=\{a_3\}$ and $S_5=\{a_1, a_4\}$. Thus, $N=4$ and $M=5$.
The minimal vertex cover is given by $S_2$
and $S_3$. The vertex set $Y$ is fully connected to $X$.
In the optimal \FTBFS\ $T^*$, $Y$ is required to be connected to the $x_j$ nodes that corresponds to the sets appearing in the optimal cover. For example, $y_\ell$
is connected to $x_2$ and $x_3$ which ``covers" the $Z$ nodes.
The red edges are necessary upon the fault of
$e_3$. All edges described except for the $(x_j,y_\ell)$ edges are required in any
 \FTBFS\ tree. }
\end{center}
\end{figure}

\paragraph{The Transformation.}
Given a Set-Cover instance $\langle U, \Set \rangle$,
we construct a Minimum \FTBFS\ instance $\mathcal{I}(U, \Set)=(G,s)$
as follows. Let $X=\{x_1, \ldots, x_M\}$ (resp.,  $Z=\{z_1, \ldots, z_N\}$)
be the vertex set corresponding to the collection of sets $\Set$
(resp., elements of $U$). Let $B_{XZ}=(X, Z, E_{XZ})$ be the bipartite
graph corresponding to the input $\langle U, \Set \rangle$, where
$E_{XZ}=\{(x_j, z_i) ~\mid ~u_i \in S_j,~ j \in \{1, \ldots ,M\} \mbox{~and~} i \in \{1, \ldots, N\}\}$.
Embed the bipartite graph $B_{XZ}$ in $G$ in the following manner.
Construct a length-$(N+1)$ path $P=[s=p_0, p_1 \ldots, p_N, p_{N+1}]$,
connect a vertex $v'$ to $p_{N}$ and connect a set of vertices $Y=\{y_1, \ldots, y_R\}$ for  $R=O((M N)^3)$ to the vertex $p_{N+1}$ by the edges of $E_{pY}=\{(p_{N+1},y_i) ~\mid~ i\in \{1, \ldots, R\}\}$.
Connect these vertices to the bipartite graph $B_{XZ}$ as follows.
For every $i\in \{1, \ldots, N\}$, connect the node $p_{i-1}$ of $P$ to the node $z_i$ of $Z$ by a path $Q_i=[p_{i-1}=q^i_0, \ldots, q^i_{t_i}=z_i]$ where $t_i=|Q_i|=6+2(N-i)$.
Thus the paths $Q_i$ are monotonely decreasing and vertex disjoint.
In addition, connect the vertices $v'$ and $p_{N+1}$ to every vertex of $X$, adding the edge sets $E_{vX}=\{(v', x_i) \mid x_i \in X\}$ and $E_{pX}=\{(p_{N+1}, x_j) \mid x_j \in X\}$.
Finally, construct a complete bipartite graph $B_{XY}=(X, Y, E_{XY})$ where
$E_{XY}=\{(y_\ell, x_j) \mid x_j \in X, y_\ell \in Y\}$.
This completes the description of $G$. For illustration, see
Fig. \ref{fig:reducbfs}.
Overall,
$$V(G)=X \cup Z \cup V(P) \cup \bigcup_{i=1}^N V(Q_i) \cup \{v'\} \cup Y,$$
and
$$E(G)=E_{XZ} \cup E(P) \cup \bigcup_{i=1}^N E(Q_i) \cup \{(p_{N}, v')\} \cup E_{pY} \cup E_{vX} \cup E_{pX} \cup E_{XY}.$$
Note that $|V(G)|=O(R)$ and that $|E(G)|=O(|E_{XZ}|+N^2+M R)=O(M R)$.

First, note the following.
\begin{observation}
\label{obs:reduc_triv_edges}
Upon the failure of the edge $e_i=(p_{i-1}, p_i)$, $i\in \{1, \ldots, N\}$, the following happen:\\
(a) the unique $s-z_{i}$ shortest path in $G \setminus \{e_i\}$ is given
by $\widetilde{P}_i=P[s, p_{i-1}] \circ Q_i$.\\
(b) the shortest-paths connecting $s$ and the vertices of $\{p_N, p_{N+1}, v'\} \cup X \cup Y$ disconnect and hence the replacement paths in $G \setminus \{e_i\}$ must go through the $Z$ nodes.
\end{observation}
\par We begin by observing that all edges except those of $B_{XY}$ are necessary in every \FTBFS\ tree $\widehat{T}\in \mathcal{T}(s,G)$.
Let $\widetilde{E}=E(G) \setminus E_{XY}$.
\begin{observation}
\label{obs:everyftbfs}
$\widetilde{E} \subseteq \widehat{T}$ for every
$\widehat{T} \in \mathcal{T}(s,G)$.
\end{observation}
\Proof
The edges of the paths $P$ and the edges of $E_{pY} \cup \{(p_{N}, v')\}$ are trivially part of every \FTBFS\ tree.
The edges of the path $Q_i$ are necessary, by Obs. \ref{obs:reduc_triv_edges}(a), upon the failure of $e_{i}$ for every $i \in \{1, \ldots, N\}$.
To see that the edges of $E_{vX}$ are necessary, note
that upon the failure of the edge $(p_N, p_{N+1})$ or the
edge $(p_{N+1}, x_j)$, the unique $s-x_j$ replacement path goes through $v'$ for every $j \in \{1, \ldots, M\}$.
Similarly, the edges $E_{pX}$ are necessary
upon the failure of $(p_{N}, v')$ or $(v', x_j)$.
\par It remains to consider the edges of $E_{XZ}$. Assume, towards contradiction, that there exists some $T' \in \mathcal{T}(s,G)$ that does not contain $e_{j,i}=(x_j, z_i) \in E_{XZ}$.
Note that by Obs. \ref{obs:reduc_triv_edges}(a), upon the failure of the edge $e_i=(p_{i-1},p_{i}) \in P$, the unique $s-x_j$ shortest-path in
$G \setminus \{e_i\}$ is $P'_{i}= \pi[p_0,p_{i-1}] \circ Q_{i}\circ [z_{i}, x_j]$, and all other alternatives are strictly longer.
Since $e_{j,i} \notin T'$, also $P'_i \nsubseteq T'$, and therefore
$\dist(s, x_j, G \setminus \{e_i\})< \dist(s, x_j, T' \setminus \{e_i\})$,
in contradiction to the fact that $T' \in \mathcal{T}(s,G)$.
The observation follows.
\QED
We now prove the correctness of the reduction and then consider gap-preservation.
Let $\widehat{T} \in \mathcal{T}(s,G)$ and define by $\Gamma(y_\ell,\widehat{T})=\{x_j ~\mid~ (x_j, y_\ell) \in \widehat{T}\}$
the $X$ nodes that are connected to $y_\ell$ in $\widehat{T}$, for every $y_\ell \in Y$.
Let $\kappa(\widehat{T})=\min\{|\Gamma(y_\ell,\widehat{T})| \mid y_\ell \in Y\}$.
Note that since the edges of $\widetilde{E}$ are necessary in every $\widehat{T} \in \mathcal{T}(s,G)$ it follows that
\begin{equation}
\label{eq:costset}
|E(\widehat{T})|\geq |\widetilde{E}|+\kappa(\widehat{T})\cdot R~.
\end{equation}

\begin{lemma}
\label{cl:reduc_a}
If $\widehat{T} \in \mathcal{T}(s,G)$ then
there exists a Set-Cover for $\langle U, \Set \rangle$
of size at most $\kappa(\widehat{T})$.
\end{lemma}
\Proof
Consider $\widehat{T} \in \mathcal{T}(s,G)$ and let $y_\ell \in Y$ be such that
$|\Gamma(y_\ell,\widehat{T})|=\kappa(\widehat{T})$.
A cover $\Set'$ for $U$ for size $\kappa(\widehat{T})$ is constructed as follows.
Let $\Set'=\{S_j ~\mid~ x_j \in  \Gamma(y_\ell,\widehat{T})\}$.
By definition, $|\Set'| = \kappa(\widehat{T})$.
We now claim that it is a cover for $U$.
Assume, towards contradiction, that there exists some $u_i \in U$
not covered by $\Set'$. Consider the graph $G'=G \setminus \{e_i\}$ where $e_i=(p_{i-1}, p_{i})$. Recall that by Obs. \ref{obs:reduc_triv_edges}(a), $\widetilde{P}_k=P[s, p_{k-1}] \circ Q_k$ is the $s-z_k$ path in $G \setminus \{e_k\}$. Note that $\widetilde{P}_k \nsubseteq G'$ for every $k>i$
and $|\widetilde{P}_k|>|\widetilde{P}_i|$ for every $k<i$.
Hence denoting the set of neighbors of $z_i$ in $X$ by
$\Gamma(z_i)=\{x_j \mid (z_i,x_j) \in E_{XZ}\}$, by Obs. \ref{obs:reduc_triv_edges}(b),
the unique $s-x_j$ shortest-path, for every $x_j \in \Gamma(z_i)$ such that $(z_i, x_j) \in E_{XY}$, is given by $P'_j=\widetilde{P}_i \circ (z_i, x_j)$.
Therefore the $s-y_\ell$ shortest-paths in $G'$ are all given by $P'_j \circ (x_j, y_\ell)$, for every $x_j \in \Gamma(z_i)$. But since $(x_j,y_\ell) \notin \widehat{T}$ for every $x_j \in \Gamma(z_i)$, we have that $\dist(s, y_\ell, G')<\dist(s, y_\ell, \widehat{T}\setminus \{e_i\})$, in contradiction to the fact that $\widehat{T} \in \mathcal{T}(s,G)$.
\QED
\begin{lemma}
\label{cl:reduc_b}
If there exists a Set-Cover of size $\kappa$ then
$\Cost^*(s, G) \leq |\widetilde{E}|+\kappa \cdot R$.
\end{lemma}
\Proof
Given a cover $\Set' \subseteq \Set$, $|\Set'|=\kappa$, construct a \FTBFS\ tree $\widehat{T} \in \mathcal{T}(s,G)$ with $|\widetilde{E}|+\kappa \cdot R$ edges as follows. Add $\widetilde{E}$ to $\widehat{T}$. In addition, for every $S_j \in \Set'$, add the edge $(y_\ell, x_j)$ to $\widehat{T}$ for every $y_\ell \in Y$. Clearly, $|E(\widehat{T})|=|\widetilde{E}|+\kappa \cdot R$. It remains to show that $\widehat{T} \in \mathcal{T}(s,G)$.
Note that there is no $s-u$ replacement path that uses any $y_\ell \in Y$ as a relay, for any $u \in V(G)$ and $y_\ell \in Y$; this holds as $X$ is connected by two alternative shortest-paths to both $p_{N+1}$ and to $v'$ and the path through $y_\ell$ is strictly longer. In addition, if the edge $e \in \{(p_{N}, p_{N+1}),(p_{N+1}, y_{\ell})\}$ fails, then the $s-y_{\ell}$ shortest path in $G \setminus \{e\}$ goes through any neighbor $x_j$ of $y_{\ell}$. Since each $y_{\ell}$ has at least one $X$ node neighbor in $\widehat{T}$, it holds that $\dist(s, y_{\ell},\widehat{T} \setminus \{e\})=\dist(s, y_{\ell},G \setminus \{e\})$.
\par Since the only missing edges of $\widehat{T}$, namely, $E(G)\setminus E(\widehat{T})$, are the edges of $E_{XY}$, it follows that
it remains to check the edges $e_i=(v_{i-1}, v_i)$ for every $i \in \{1, \ldots, N\}$. Let $S_j \in \Set'$ such that $u_i \in S_j$. Since $\Set'$ is a cover, such $S_j$ exists. Hence, the optimal $s-y_\ell$ replacement path in
$G\setminus \{e_i\}$, which is by Obs. \ref{obs:reduc_triv_edges}(b), $P'=\widetilde{P}_i \circ (z_i,x_j) \circ (x_j,y_\ell)$, exists in $\widehat{T} \setminus \{e_i\}$ for every $y_\ell \in Y$.
It follows that $\widehat{T} \in \mathcal{T}(s,G)$, hence $\Cost^*(s, G) \leq |E(\widehat{T})|=|\widetilde{E}|+\kappa \cdot R$. The lemma follows.
\QED
Let $\kappa^*$ be the cost of the optimal Set-Cover for the instance $\langle U, \Set \rangle$. We have the following.
\begin{corollary}
\label{cor:reduc_opt}
$\Cost^*(s,G)=|\widetilde{E}|+\kappa^* \cdot R$.
\end{corollary}
\Proof
Let $T^* \in \mathcal{T}(s,G)$ be such that $|E(T^*)|=\Cost^*(s,G)$.
It then holds that
$$|\widetilde{E}|+\kappa(T^*) \cdot R \leq |E(T^*)|=\Cost^*(s,G)\leq |\widetilde{E}|+\kappa^* \cdot R,$$
where the first inequality holds by Eq. (\ref{eq:costset}) and the second inequality follows by Lemma \ref{cl:reduc_b}. Hence, $\kappa(T^*) \leq \kappa^*$.
Since by Lemma \ref{cl:reduc_a},
there exists a cover of size $\kappa(T^*)$, we have that  $\kappa^*\leq \kappa(T^*)$.
It follows that $\kappa^*=\kappa(T^*)$ and $\Cost^*(s,G)=|\widetilde{E}|+\kappa^* \cdot R$ as desired.
\QED
We now show that the reduction is gap-preserving.
Assume that there exists an $\alpha$ approximation algorithm $\mathcal{A}$ for the Minimum \FTBFS\ problem. Then applying our transformation to an instance $\mathcal{I}(U, \Set)=(G,s)$ would result in an \FTBFS\ tree $\widehat{T} \in \mathcal{T}(s,G)$ such that
\begin{eqnarray*}
|\widetilde{E}|+\kappa(\widehat{T})\cdot R < |E(\widehat{T})| \leq \alpha(|\widetilde{E}|+\kappa^* \cdot R)\leq 3\alpha \cdot \kappa^* \cdot R~,
\end{eqnarray*}
where the first inequality follows by Eq. (\ref{eq:costset}), the second by the approximation guarantee of $\mathcal{A}$ and by Cor. \ref{cor:reduc_opt}, and the third inequality follows by the fact that $|\widetilde{E}| \leq 2R$.
By Lemma \ref{cl:reduc_a}, a cover of size $\kappa(\widehat{T}) \leq 3\alpha \kappa^*$ can be constructed given $\widehat{T}$, which results in a $3\alpha$ approximation to the Set-Cover instance. As the Set-Cover problem is inapproximable within a factor of $(1-o(1)) \ln n$, under an appropriate complexity assumption \cite{Feige98}, we get that the Minimum \FTBFS\ problem is inapproximable within a factor of $c \cdot \log N$ for some constant $c>0$.
This complete the proof of Thm. \ref{thm:hardness}.

\section{$O(\log n)$-Approximation for FT-MBFS Trees}
\label{sec:approx}
\begin{figure}[h!]
\begin{center}
\includegraphics[scale=0.36]{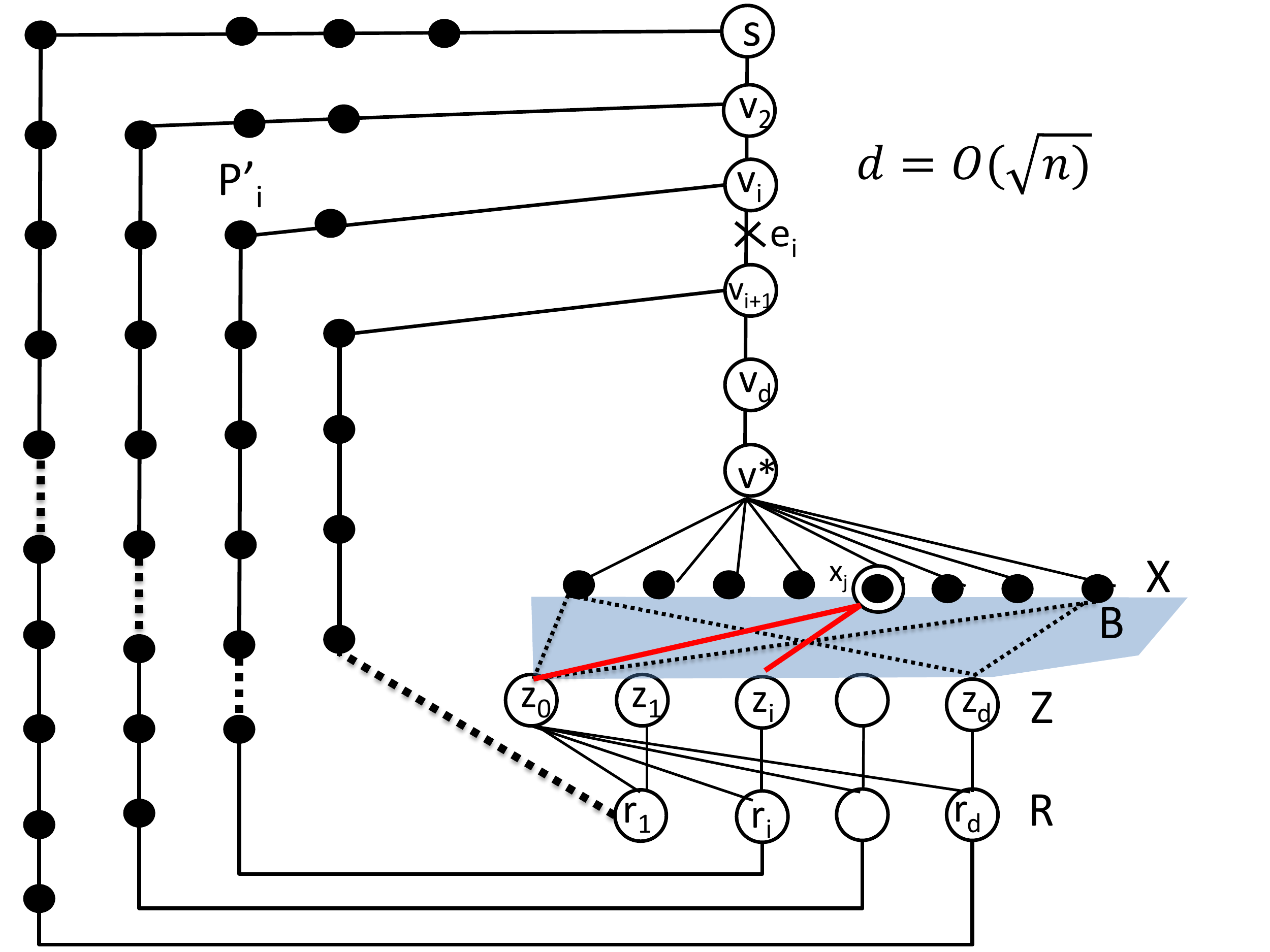}
\caption{Bad example for the algorithm of Sec. \ref{sec:upper}.
The weights of the $z_0$ edges are larger than those of the other edges. Thus, the entire complete bipartite graph $B(X,Z \setminus \{z_0\})$ of
size $\Omega(n^{3/2})$ is included in the resulting \FTBFS\ tree
$\widehat{T} \in \mathcal{T}(s,G)$ returned by the algorithm.
However, an \FTBFS\ tree $T^*$ of $O(n)$ edges
can be given by including the edges of $(z_0,x_i)$ for every $x_i \in X$.
The red edges are two optional edges necessary upon the failure of $e_i$. Adding the edge $(x_j,z_0)$ is better, yet the algorithm of Sec. \ref{sec:upper} adds $(x_j,z_i)$ to $\widehat{T}$ for every $x_j \in X$.
\label{fig:badexamp}}
\end{center}
\end{figure}

In Sec. \ref{subsec:1edge_alg}, we presented an algorithm that for every graph $G$ and source $s$ constructs an \FTBFS\ tree
$\widehat{T} \in \mathcal{T}(s,G)$ with $O(n^{3/2})$ edges.
In Sec. \ref{sc:lb_single},
we showed that there exist graphs $G$ and $s \in V(G)$ for which $\Cost^*(s, G)=\Omega(n^{3/2})$, establishing tightness of our algorithm in the worst-case.
Yet, there are also inputs $(G',s')$ for which the algorithm of Sec. \ref{sec:upper}, as well as algorithms based on the analysis of \cite{GW12} and \cite{RTREP05}, might still produce an \FTBFS\
$\widehat{T} \in \mathcal{T}(s',G')$ which is denser by a factor of
$\Omega(\sqrt{n})$ than the size of the optimal \FTBFS\ tree, i.e., such that $|E(\widehat{T})|\geq \Omega(\sqrt{n}) \cdot \Cost^*(s', G')$.
For an illustration of such a case consider the graph $G'=(V,E)$ which is a modification of the graph $G$ described in
Sec. \ref{sc:lb_single}.
The modifications are as follows.
First, add a node $z_0$ to $Z$ and connect it to every $x_i \in X$.
Replace the last edge $e'_i=\LastE(P_i)$ of the $v_i-z_i$ path $P_i$ by a
vertex $r_i$ that is connected to the endpoints of the edge $e'_i$
for every $i \in \{1, \ldots, d\}$.
Let $P'_i$ be the $s-z_i$ modified path where $\LastE(P'_i)=(r_i,z_i)$.
Finally, connect the node $z_0$ to all nodes $r_i$ for every $i \in \{1, \ldots, d\}$.
See Fig. \ref{fig:badexamp} for illustration.

Observe that whereas $\Cost^*(s, G)=\Omega(n^{3/2})$,
the modified $G'$ has $\Cost^*(s, G')=O(n)$, as the edges of the
complete bipartite graph $B$ that are required in
every $\widehat{T} \in \mathcal{T}(s,G)$
are no longer required in every $T' \in \mathcal{T}(s,G')$;
it is sufficient to connect the nodes of $X$ to $z_0$ only,
and by that ``save" the $\Omega(n^{3/2})$ edges of $B$ in $T'$.
Nevertheless, as we show next, for certain weight assignments the algorithm of
Sec. \ref{sec:upper} constructs an \FTBFS\ tree $\widehat{T}$ of size
$O(n^{3/2})$.
Specifically, let $W$ be such that each of the edges of
$$E' ~=~
\{(z_0,r_i) ~\mid~ i\in \{1, \ldots, d\}\} \cup \{(z_0,x_i) ~\mid~ x_i \in X\}$$
is assigned a weight which is strictly larger than the weights of the other edges. That is, $W(e_k)>W(e_\ell)$ for every $e_k \in E'$ and
$e_{\ell} \in E(G')\setminus E'$.
Note that for every edge $e_i=(v_{i}, v_{i+1})\in \pi$, $i \in \{1, \ldots,d\}$,
there are two alternative $s-x_j$ replacement paths of the same length, namely,
$Q_{i,j}=\pi[s, v_i] \circ P'_i \circ (z_i, x_j)$ that goes through $z_i$
and $\widehat{Q}_{i,j}=\pi[s, v_i] \circ P'_i[s, r_i] \circ (r_i,z_0) \circ (z_0,x_i)$
that goes through $z_0$. Although $|Q_{i,j}|=|\widehat{Q}_{i,j}|$,
the weight assignment implies that
$\Cost(Q_{i,j})<\Cost(\widehat{Q}_{i,j})$
and hence $\widehat{Q}_{i,j} \notin SP(s, x_j, G \setminus \{e_i\},W)$ for
every $i \in \{1, \ldots,d\}$ and every $x_j \in X$.
Therefore, $E(B) \subseteq  \widehat{T}$, for every \FTBFS\ tree
$\widehat{T}$ computed by the
algorithm of Sec. \ref{sec:upper} with the weight assignment $W$.
Hence $|E(\widehat{T})|=\Theta(n^{3/2})$ while $\Cost^*(s, G')=O(n)$.

Clearly, a universally optimal algorithm is unlikely given the hardness
of approximation result of Thm. \ref{thm:hardness}.
Yet the gap can be narrowed down.
The goal of this section is to present an $O(\log n)$ approximation algorithm
for the Minimum \FTBFS\ Problem (hence also to its special case,
the Minimum \FTBFS\ Problem, where $|S|=1$).

To establish this result, we first describe the algorithm and
then bound the number of edges.
Let $\ApproxSetCover(\Set,U)$ be an $O(\log n)$
approximation algorithm for the Set-Cover problem,
which given a collection of sets $\Set=\{S_1, \ldots, S_{M}\}$
that covers a universe $U=\{u_1, \ldots, u_N\}$ of size $N$, returns a cover
$\Set' \subseteq \Set$ that is larger by at most
$O(\log N)$ than any other $\Set'' \subseteq \Set$ that covers $U$ (cf. \cite{Vazirani97}).

\paragraph{The Algorithm}
Starting with $\widehat{T}=\emptyset$, the algorithm adds edges to $\widehat{T}$ until it becomes an \FTMBFS\ tree.
\par Set an arbitrary order on the vertices
$V(G)=\{v_1, \ldots, v_{n}\}$ and on the edges
$E^+=E(G) \cup \{e_0\}=\{e_0, \ldots, e_m\}$ where $e_0$ is a new fictitious edge whose role will be explained later on.
For every node $v_i \in V$, define
$$U_i=\{ \langle s_k, e_j \rangle ~\mid s_k \in S \setminus \{v_i\},e_j \in E^{+}\}.$$
The algorithm consists of $n$ rounds, where in round $i$
it considers $v_i$. Let $\Gamma(v_i, G)=\{u_1, \ldots, u_{d_i}\}$
be the set of neighbors of $v_i$ in some arbitrary order, where $d_i=\deg(v_i,G)$. For every neighbor $u_j$, define a set $S_{i,j} \subseteq U_i$ containing certain source-edge pairs $\langle s_k, e_\ell \rangle \in U_i$. Informally, a set $S_{i,j}$ contains the pair  $\langle s_k, e_\ell \rangle$ iff there exists an $s_k-v_i$ shortest path in $G \setminus \{e_\ell\}$ that goes through the neighbor $u_j$ of $v_i$.
Note that $S_{i,j}$ contains the pair  $\langle s_k, e_0 \rangle$
iff there exists an $s_k-v_i$ shortest-path in $G \setminus \{e_0\}=G$ that goes through $u_j$. I.e., the fictitious edge $e_0$ is meant to capture the case where no fault occurs, and thus we take care of true shortest-paths in $G$. Formally, every pair $\langle s_k, e_\ell \rangle \in U_i$ is included
in every set $S_{i,j}$ satisfying that
\begin{equation}
\label{eq:intset}
\dist(s_k, u_j, G \setminus \{e_\ell\})=\dist(s_k, v_i, G \setminus \{e_\ell\})-1.
\end{equation}
Let $\Set_i=\{S_{i,1}, \ldots, S_{i,d_i}\}$.
The edges of $v_i$ that are added to $\widehat{T}$ in round $i$ are now selected by using algorithm $\ApproxSetCover$ to generate an approximate solution for the set cover problem on the collection $\Set=\{S_{i,j}~\mid~ u_j \in \Gamma(v_i, G)\}$.
Let $\Set'_i=\ApproxSetCover(\Set_i,U_i)$.
For every $S_{i,j}\in \Set'_i$, add the edge
$(u_j,v_i)$ to $\widehat{T}$.
We now turn to prove the correctness of this algorithm
and establish Thm. \ref{thm:edgeonef-approx}.

\begin{figure}[htb!]
\begin{center}
\includegraphics[scale=0.7]{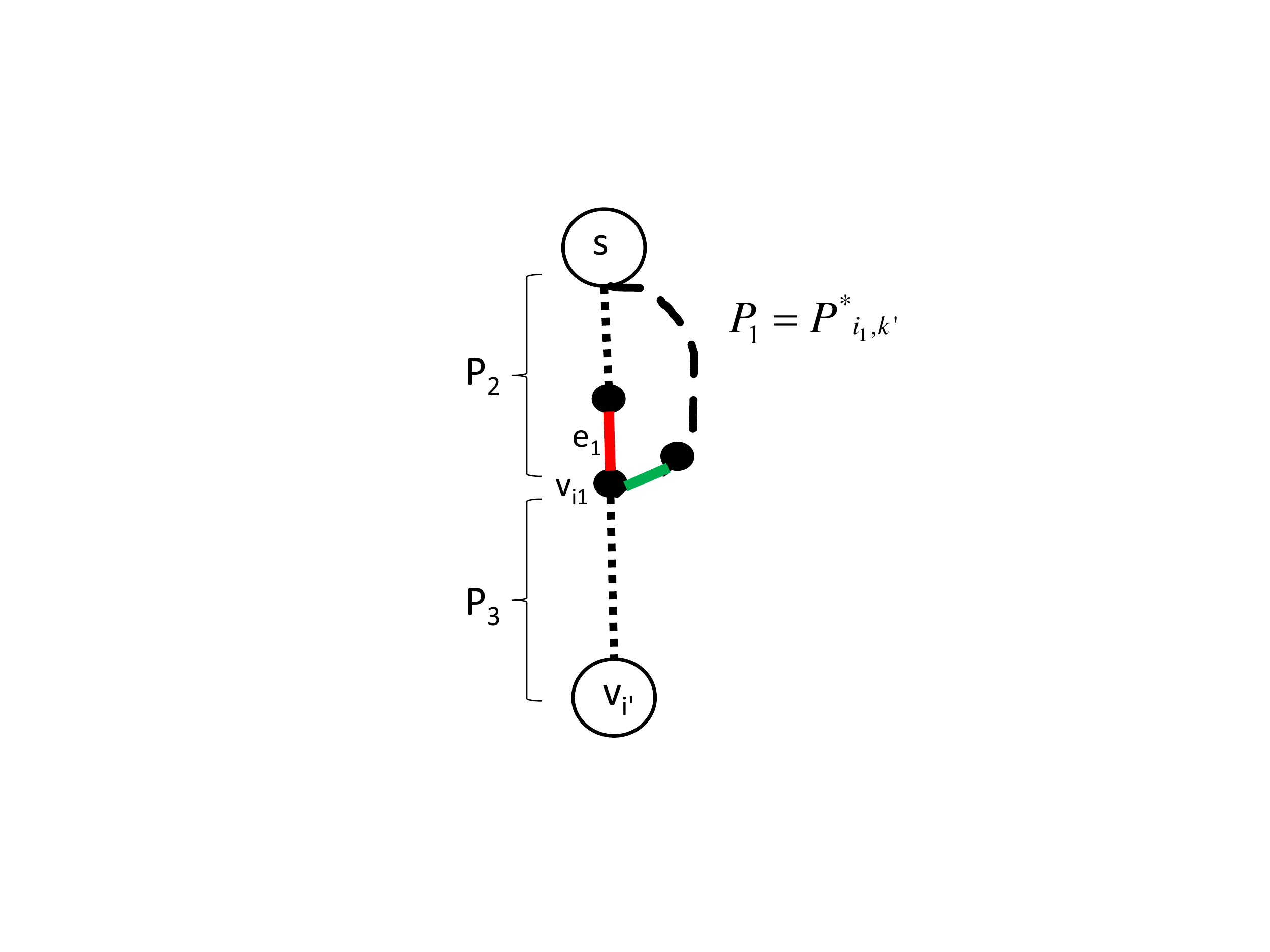}
\caption{\label{fig:guidedpaths}
Red solid lines correspond to new edges.
The ``deepest missing edge'' for $(i',k')$, edge $e_1$, is the shallowest
such edge over all bad pairs in $BP$. Yet the pair $(i_1,k')$ is bad too.
As the last (green) edge of $P_1$ is included in the \FTMBFS\ tree,
and since $P_1$ and $P_2$ are of the same length, it follows that $P_1$
has a shallower ``deepest missing edge''.}
\end{center}
\end{figure}

\paragraph{Analysis}
We first show that algorithm constructs an \FTMBFS\
$\widehat{T} \in \mathcal{T}(S,G)$ and then bound its size.
\begin{lemma}
\label{lem:correct}
$\widehat{T} \in \mathcal{T}(S,G)$.
\end{lemma}
\Proof
Assume, towards contradiction, that $\widehat{T} \notin \mathcal{T}(S,G)$.
Let $s \in S$ be some source node such that $\widehat{T} \notin \mathcal{T}(s,G)$ is not an \FTBFS\ tree with respect to $s$. By the assumption, such $s$ exists. Let
$$BP=\{(i,k) \mid v_i \in V, e_k \in E^{+} \mbox{~and~}
\dist(s,v_i, \widehat{T} \setminus \{e_k\}) >  \dist(s,v_i,G \setminus \{e_k\})\}$$
be the set of ``bad pairs," namely, vertex-edge pairs $(i,k)$ for which the $s-v_i$ shortest path distance in $\widehat{T} \setminus \{e_k\}$ is greater than that in $G\setminus \{e_k\}$.
(By the assumption that $\widehat{T} \notin \mathcal{T}(s,G)$, it holds that $BP\ne \emptyset$.)
For every vertex-edge pair $(i,k)$, where $v_i \in V \setminus \{s\}$ and $e_k \in E^{+}$,
define an $s-v_i$ shortest-path $P^*_{i,k}$ in $G \setminus \{e_k\}$ in the following manner.
Let $u_j \in \Gamma(v_i, G)$ be such that the pair $\langle s, e_k \rangle \in S_{i,j}$
is covered by the set $S_{i,j}$ of $u_j$ and $S_{i,j} \in \Set'_i$ is included in the cover returned by the algorithm $\ApproxSetCover$ in round $i$. Thus, $(u_j,v_i)\in \widehat{T}$ and
$\dist(s, u_j, G \setminus \{e_k\})=\dist(s, v_i, G \setminus \{e_k\})-1$.
Let $P' \in SP(s, u_j, G \setminus \{e_k\})$ and define
$$P^*_{i,k}=P' \circ (u_j,v_i).$$
By definition, $|P^*_{i,k}|=\dist(s, v_i, G \setminus \{e_k\})$ and by construction, $\LastE(P^*_{i,k}) \in \widehat{T}$.
Define
$BE(i,k)=P^{*}_{i,k} \setminus E(\widehat{T})$ to be the set of ``bad edges,''
namely, the set of $P^{*}_{i,k}$ edges that are missing in $\widehat{T}$.
By definition, $BE(i,k) \neq \emptyset$ for every bad pair $(i,k) \in BP$.
Let $d(i,k)=\max_{e \in BE(i,k)}\{\dist(s,e,P^{*}_{i,k})\}$ be the maximal depth
of a missing edge in $BE(i,k)$, and let $DM(i,k)$ denote that ``deepest
missing edge'' for $(i,k)$, i.e., the edge $e$ on $P^{*}_{i,k}$ satisfying
$d(i,k) = \dist(s,e,P^{*}_{i,k})$.
Finally, let $(i',k') \in BP$ be the pair that minimizes $d(i,k)$,
and let $e_1=(v_{\ell_1}, v_{i_1}) \in BE(i',k')$ be
the deepest missing edge on $P^{*}_{i',k'}$, namely, $e_1=DM(i',k')$.
Note that $e_1$ is the {\em shallowest} ``deepest missing edge''
over all bad pairs $(i,k) \in BP$.
Let $P_1=P^{*}_{i_1,k'}$, $P_2=P^{*}_{i',k'}[s,v_{i_1}]$ and
$P_3=P^{*}_{i',k'}[v_{i_1}, v_{i'}]$;
see Fig. \ref{fig:guidedpaths} for illustration.
Note that since $(i',k') \in BP$, it follows that also $(i_1, k') \in BP$.
(Otherwise, if $(i_1, k') \notin BP$, then any $s-v_{i_1}$ shortest-path
$P' \in SP(s, v_{i_1}, \widehat{T} \setminus \{e_{k'}\})$
, where $|P'|=|P^*_{i_1,k'}|$, can be appended to $P_3$ resulting in
$P''=P' \circ P_3$
such that (1) $P'' \subseteq \widehat{T}\setminus \{e_{k'}\}$ and (2)
$|P''|=|P'|+|P_3|=|P_2|+|P_3|=|P^{*}_{i',k'}|$, contradicting the fact that
$(i',k') \in BP$.)
Thus we conclude that $(i_1, k') \in BP$.
Finally, note that $\LastE(P_1) \in \widehat{T}$ by definition, and therefore
the deepest missing edge of $(i,k)$ must be shallower, i.e.,
$d(i_1,k')<d(i',k')$. However, this is in contradiction to our choice
of the pair $(i',k')$. The lemma follows.
\QED

Let $W:E(G) \to \mathbb{R}_{>0}$ be the weight assignment that guarantees the uniqueness of shortest-paths. Note that the algorithm did not use $W$ in the computation of the shortest-paths. For every node $v_i$, let
$\Gamma(v_i,G)=\{u_{1}, \ldots, u_{d_i}\}$ be its ordered neighbor set
as considered by the algorithm.
For every \FTMBFS\ tree
$\widetilde{T}\in \mathcal{T}(S,G)$, $v_i \in V, e_\ell \in E^{+}$ and $s_k \in S$, let $\widetilde{P}_i(s_k,e_\ell) \in SP(s_k, v_i, \widetilde{T} \setminus \{e_\ell\},W)$ be an $s_k-v_i$ shortest-path in $\widetilde{T} \setminus \{e_\ell\}$. Let
$$A_i(\widetilde{T})=\{\LastE(\widetilde{P}_i(s_k,e_\ell)) ~\mid~ e_\ell \in E^{+}, s_k \in S \setminus \{v_i\} \}$$
be the edges of $v_i$ that appear as last edges in the shortest-paths and replacement paths from $S$ to $v_i$ in $\widetilde{T}$.
Define
$$\Set_i(\widetilde{T})=\{S_{i,j} ~\mid~ (u_j, v_i) \in A_i(\widetilde{T})\}.$$
We then have that
\begin{equation}
\label{eq:set_size}
|\Set_i(\widetilde{T})|=|A_i(\widetilde{T})|~.
\end{equation}
The correctness of the algorithm (see Lemma \ref{lem:correct})
established that if a subgraph $\widetilde{T} \subseteq G$ satisfies that
$\Set_i(\widetilde{T})$ is a cover of
$U_i$ for every $v_i \in V$, then $\widetilde{T} \in \mathcal{T}(S,G)$.
We now turn to show the reverse direction.
\begin{lemma}
\label{lem:cover_tree}
For every $\widetilde{T} \in \mathcal{T}(S,G)$, the collection $\Set_i(\widetilde{T})$ is a cover of $U_i$, namely,
$\bigcup_{S_{i,j} \in \Set_i(\widetilde{T})}S_{i,j}=U_i, \mbox{~~for every~~} v_i \in V$.
\end{lemma}
\Proof
Assume, towards contradiction, that there exists an \FTMBFS\ tree
$\widetilde{T} \in \mathcal{T}(S,G)$ and a vertex $v_i \in V$
whose corresponding collection of sets $\Set_i(\widetilde{T})$
does not cover $U_i$.
Hence there exists at least one uncovered pair $\langle s_k, e_\ell \rangle \in U_i$, i.e.,
\begin{equation}
\label{eq:notcovered}
\langle s_k, e_\ell \rangle \in U_i \setminus\bigcup_{S_{i,j} \in \Set_i(\widetilde{T})}S_{i,j}~.
\end{equation}
By definition $s_k \neq v_i$. We next claim that $\widetilde{T}$ does not contain an optimal
$s_k-v_i$ path when the edge $e_\ell$ fails,
contradicting the fact that $\widetilde{T} \in \mathcal{T}(S,G)$.
That is, we show that
$$\dist(s_k, v_i, \widetilde{T} \setminus \{e_\ell\})>\dist(s_k, v_i, G \setminus \{e_\ell\}).$$
Towards contradiction, assume otherwise, and let
$(u_j, v_i)=\LastE(P^*_{i,\ell})$
where $P^*_{i,\ell} \in SP(s_k, v_i, \widetilde{T} \setminus \{e_\ell\},W)$,
hence $(u_j, v_i) \in A_i(\widetilde{T})$
and $S_{i,j} \in \Set_i(\widetilde{T})$.
By the contradictory assumption, $|P^*_{i,\ell}|=\dist(s_k, v_i, G \setminus \{e_\ell\})$
and hence $\dist(s_k, u_j, G \setminus \{e_\ell\})=\dist(s_k, v_i, G \setminus \{e_\ell\})-1$.
This implies that $\langle s_k,e_\ell \rangle \in S_{i,j} \in \Set_i(\widetilde{T})$, in contradiction to Eq. (\ref{eq:notcovered}),
stating that $\langle s_k,e_\ell \rangle$ is not covered by $\Set_i(\widetilde{T})$.
The lemma follows.
\QED
We now turn to bound that number of edges in $\widehat{T}$.
\begin{lemma}
\label{lem:numbersize}
$|E(\widehat{T})| \leq O(\log n)\cdot \Cost^*(S, G)$.
\end{lemma}
\Proof
Let $\delta=c\log n$ be the approximation
ratio guarantee of $\ApproxSetCover$.
For ease of notation, let
$O_i=A_i(T^*)$ for every $v_i \in V$.
Let $\Set_i=\{S_{i,1}, \ldots, S_{i,d_i}\}$ be the collection of
$v_i$ sets considered at round $i$ where $S_{i,j}\subseteq U_i$
is the set of the neighbor $u_j \in \Gamma(v_i,G)$ computed according to
Eq. (\ref{eq:intset}).
\par Let $\Set'_i=\ApproxSetCover(\mathcal{S}_i,U_i)$
be the cover returned by the algorithm and define
$A_i=\{(u_j, v_i) ~\mid~ S_{i,j} \in \Set'_i\}$
as the collection of edges whose corresponding sets are included in
$\mathcal{S}'_i$.
Thus, by Eq. (\ref{eq:set_size}),
$|O_i|=|\Set_i(T^*)|$ and $|A_i|=|\Set'_i|$ for every $v_i \in V$.
\begin{observation}
\label{cl:a_i}
$|A_i| \leq \delta |O_i|$ for every $v_i \in V \setminus \{s\}$.
\end{observation}
\Proof
Assume, towards contradiction, that there exists some $i$ such that
$|A_i|>\delta |O_i|$. Then by Eq. (\ref{eq:set_size})
and by the approximation guarantee of \ApproxSetCover\, where in
particular $|\Set_i(\widetilde{T})| \leq \delta |\Set''_i|$
for every $\Set''_i \subseteq  \Set_i$ that covers $U_i$, it follows that $\Set_i(T^*)$ is not a cover of $U_i$.
Consequently, it follows by Lemma \ref{lem:cover_tree}
that $T^* \notin \mathcal{T}(S,G)$, contradiction.
The observation follows.
\QED
Since $\bigcup A_i$ contains precisely the edges that are added by the algorithm to the constructed \FTMBFS\ tree $\widehat{T}$, we have that
\begin{eqnarray*}
|E(\widehat{T})|&\leq& \sum_i |A_i| \leq
\delta \sum_i |O_i| \leq 2\delta \cdot \Cost^*(S, G)~,
\end{eqnarray*}
where the second inequality follows by Obs. \ref{cl:a_i} and
the third by the fact that $|E(T^*)| \geq \sum_i |O_i|/2$ (as every edge in $\bigcup_{v_i \in V} O_i$ can be counted at most twice, by both its endpoints). The lemma follows.
\QED

The following theorem is established.

\begin{theorem}
\label{thm:edgeonef-approx}
There exists a polynomial time algorithm that for every $n$-vertex graph $G$
and source node set $S \subseteq V$ constructs an \FTMBFS\ tree
$\widehat{T} \in \mathcal{T}(S,G)$
such that $|E(\widehat{T})|\leq O(\log n)\cdot \Cost^*(S, G)$.
\end{theorem}

%

\bigskip
\paragraph{Acknowledgment}
We are grateful to Gilad Braunschvig, Alon Brutzkus, Adam Sealfon
and Oren Weimann for helpful discussions.

\clearpage


\begin{thebibliography}{10}



\bibitem{ABLP-89:stoc}
B. Awerbuch, A. {Bar-Noy}, N. Linial, and D. Peleg.
\newblock Compact distributed data structures for adaptive network routing.
\newblock
In {\em Proc. 21st ACM Symp. on Theory of Computing},
230--240, 1989.






\bibitem{ACGP10}
I. Abraham, S. Chechik, C. Gavoille and D. Peleg.
\newblock Forbidden-Set Distance Labels for Graphs of Bounded Doubling
Dimension.
\newblock
In {\em Proc. 29th ACM Symp. on Principles of Distributed Computing},
2010, 192--200.


\bibitem{BS06}
S.~Baswana and S.~Sen.
\newblock Approximate distance oracles for unweighted graphs in expected
  {$O(n^2)$} time.
\newblock {\em ACM Trans. Algorithms}, 2(4):557--577, 2006.


\bibitem{BK09}
A.~Bernstein and D.~Karger.
\newblock A nearly optimal oracle for avoiding failed vertices and edges.
\newblock
In {\em Proc. 41st ACM Symp. on Theory of Computing},
101--110, 2009.


\bibitem{CLPR09-do}
S.~Chechik, M.~Langberg, D.~Peleg, and L.~Roditty.
\newblock $f$-sensitivity distance oracles and routing schemes.
\newblock {\em Algorithmica}, 861--882, 2012.

\bibitem{CLPR09-span}
S.~Chechik, M.~Langberg, D.~Peleg, and L.~Roditty.
\newblock Fault-tolerant spanners for general graphs.
\newblock
In {\em Proc. 41st ACM Symp. on Theory of computing},
435--444, 2009.

\bibitem{S12}
S. Chechik.
\newblock Fault-Tolerant Compact Routing Schemes for General Graphs.
\newblock
In {\em Proc. 38th Int. Colloq. on Automata, Languages \& Prog.},
101--112, 2011.



\bibitem{CT07}
B.~Courcelle and A.~Twigg.
\newblock Compact forbidden-set routing.
\newblock
In {\em Proc. 24th Symp. on Theoretical Aspects of Computer Science},
37--48, 2007.

\bibitem{CZ03}
A.~Czumaj and H.~Zhao.
\newblock Fault-tolerant geometric spanners.
\newblock {\em Discrete \& Computational Geometry}, 32, 2003.

\bibitem{DTCR08}
C.~Demetrescu, M.~Thorup, R.~Chowdhury, and V.~Ramachandran.
\newblock Oracles for distances avoiding a failed node or link.
\newblock {\em SIAM J. Computing}, 37:1299--1318, 2008.

\bibitem{DK11}
M. Dinitz and R. Krauthgamer.
\newblock Fault-tolerant spanners: better and simpler.
\newblock
In {\em Proc. ACM Symp. on Principles of Distributed Computing},
2011, 169-178.


\bibitem{DP09}
R.~Duan and S.~Pettie.
\newblock Dual-failure distance and connectivity oracles.
\newblock
In {\em Proc. 20th ACM-SIAM Symp. on Discrete Algorithms},
2009.







\bibitem{Feige98}
U.~Feige.
\newblock A Threshold of ln n for Approximating Set Cover.
\newblock {\em J. ACM}, 634--652, 1998.

\bibitem{GW12}
F.~Grandoni and V.V~Williams.
\newblock Improved Distance Sensitivity Oracles via Fast Single-Source Replacement Paths.
\newblock
In {\em Proc. 53rd IEEE Symp. on Foundations of Computer Science},
2012.

\bibitem{LNS98}
C.~Levcopoulos, G.~Narasimhan, and M.~Smid.
\newblock Efficient algorithms for constructing fault-tolerant geometric
  spanners.
\newblock
In {\em Proc. 30th ACM Symp. on Theory of computing},
186--195, 1998.

\bibitem{L99}
T.~Lukovszki.
\newblock New results of fault tolerant geometric spanners.
\newblock
In {\em Proc. 6th Workshop on Algorithms and Data Structures}, London
193--204, 1999.



\bibitem{Peleg00:book}
D.~Peleg.
\newblock {\em Distributed Computing: A Locality-Sensitive Approach}.
\newblock SIAM, 2000.


\bibitem{P09}
D. Peleg.
\newblock As good as it gets: Competitive fault tolerance in network structures.
\newblock
In {\em Proc. 11th Symp. on Stabilization, Safety, and Security
of Distributed Systems}, LNCS 5873,
2009, 35--46.



\bibitem{PelegS-89}
D. Peleg and A.A. Sch\"affer.
\newblock Graph spanners.
\newblock {\em J. Graph Theory}, 13:99--116, 1989.

\bibitem{PelegU-89}
D. Peleg and J.D. Ullman.
\newblock An optimal synchronizer for the hypercube.
\newblock {\em SIAM J. Computing}, 18(2):740--747, 1989.


\bibitem{PU-89:tables}
D.~Peleg and E.~Upfal.
\newblock A trade-off between space and efficiency for routing tables.
\newblock {\em J. ACM}, 36:510--530, 1989.

\bibitem{RTZ05}
L.~Roditty, M.~Thorup, and U.~Zwick.
\newblock Deterministic constructions of approximate distance oracles and
  spanners.
\newblock
In {\em Proc. 32nd Int. Colloq. on Automata, Languages \& Prog.},
261--272, 2005.


\bibitem{RTREP05}
L.~Roditty and U.~Zwick.
\newblock Replacement paths and k simple shortest paths in unweighted directed graphs.
\newblock {\em ACM Trans. Algorithms} ,2012.


\bibitem{TZ01}
M.~Thorup and U.~Zwick.
\newblock Compact routing schemes.
\newblock
In {\em Proc. 14th ACM Symp. on Parallel Algorithms and Architecture}, Hersonissos, Crete,
1--10, 2001.

\bibitem{TZ05}
M.~Thorup and U.~Zwick.
\newblock Approximate distance oracles.
\newblock {\em J. ACM}, 52:1--24, 2005.

\bibitem{TH99}
M.~Thorup.
\newblock Undirected single-source shortest paths with positive integer weights in linear time.
\newblock {\em J. ACM}, 362--394, 1999.

\bibitem{Vazirani97}
V.~Vazirani.
\newblock Approximation Algorithms.
\newblock
{\em College of Computing, Georgia Institute of Technology},
1997.

\bibitem{WY10}
O.~Weimann and R.~Yuster.
\newblock Replacement paths via fast matrix multiplication.
\newblock
In {\em Proc. 51th IEEE Symp. on Foundations of Computer Science},
2010.


\end{thebibliography}



\end{document}